\documentclass{aa}
\usepackage{upgreek}
\usepackage{graphicx}
\usepackage[varg]{txfonts}
\usepackage{natbib}
\usepackage{float}
\usepackage{soul}
\newlength{\linwx}
\usepackage{color}
\usepackage{lipsum}

\begin{document}

\title{Diversity of planetary origins and compositions traced by stellar abundance differences of co-natal star systems}
\author{
Bertram Bitsch \inst{1},
Fan Liu \inst{2},
Haiyang S. Wang \inst{3}
}
\offprints{B. Bitsch,\\ \email{bbitsch@ucc.ie}}
\institute{
Department of Physics, University College Cork, Cork, T12 R229, Ireland
\and
National Astronomical Observatories, Chinese Academy of Sciences, Beijing 100101, China
\and
Centre for Star and Planet Formation, Globe Institute, University of Copenhagen, Øster Voldgade 5-7, 1350 Copenhagen, Denmark
}
\abstract{So called co-natal stars, wide binaries that form from the same molecular cloud, should harbor the same stellar composition. However, observations indicate stellar abundance differences in a small fraction of these co-natal systems. These stellar abundance differences have been attributed to the ingestion of planets by their host stars. We analyze a previous sample of co-natal stars to determine the composition of the planets that were ingested by their host stars. The abundance differences in these co-natal pairs come in different flavors, where systems can show a positive abundance difference of the refractory elements (Fe, Mg, Si) paired with a positive, zero, or negative abundance difference of volatiles (O, C, S). Positive trends of refractory and volatile abundance differences can be explained (at minimum) with the ingestion of a single planet by one of the stars. However, positive trends of refractory abundance differences paired with zero or negative abundance differences of oxygen and carbon imply that both stars are required to ingest planets formed at different locations (interior/exterior to the water and/or carbon grain evaporation front). Furthermore, some systems show a sulfur abundance difference that is a factor of a few lower than the abundance difference of Fe, Si, and Mg, even though sulfur is mostly accreted via the moderately volatile compound FeS. We show that this can be explained by the migration of one of the planets across the FeS evaporation front, where the planet first accretes large amounts of sulfur with the solids and then no sulfur, while still accreting the other refractory elements before being ingested by its host star. Our study shows that the planetary compositions derived from the detailed stellar abundances can explain the measured abundance differences of co-natal stars, giving a handle on the composition of exoplanets. At the same time it implies very different planetary compositions caused by the different formation location of these planets, in agreement with planet formation models that study the formation of inner super-Earths and sub-Neptunes.
}
\authorrunning{Bitsch et al.}\maketitle

\section{Introduction}

Transiting photometry \citep{2000ApJ...529L..45C} and radial velocity \citep{1995Natur.378..355M} can help to determine planetary radii and masses, respectively. If both techniques can be combined for a single target, it unlocks new information: the planetary density, which can give information about the planetary composition for low mass planets (e.g. \citealt{2019PNAS..116.9723Z, 2019MNRAS.484..712D, 2021Sci...374..330A}). However, for many of these planets, the solutions are degenerate and one can not determine the composition with certainty (e.g. \citealt{2015A&A...577A..83D, 2022MNRAS.513.5829W, 2026PSJ.....7...30S}), even though differences between rocky super-Earths and sub-Neptunes that feature small atmospheres can be made.

The planetary composition is thought to help to unlock the formation location of small planets. If precise measurements of mass and radius exist (of the order of less than 10\%), the formation locations of rocky planets could indeed be determined through detailed models (e.g. \citealt{2019MNRAS.484..712D}) that even allow to distinguish between iron rich and iron poor formation within the same planetary system. However, the data of this quality is rare, but will hopefully improve with the launch of PLATO \citep{2025ExA....59...26R}.

From a statistical perspective, the discovery of the radius valley of small exoplanets \citep{2017AJ....154..109F}, where smaller planets are presumably of rocky nature, while larger sub-Neptues are thought to be rocky cores with small gaseous envelopes, lead to the idea that maybe the formation region of these planets is in the inner protoplanetary disc (e.g. \citealt{2013ApJ...775..105O, 2017ApJ...847...29O, 2018MNRAS.479.4786V}). However, recent observations show that sub-Neptunes could harbor large fractions of water vapor (e.g. \citealt{2024arXiv240303325B}), consistent with a formation region exterior to the water ice line. Furthermore, simulations of planet formation indicate that probably a mixture between water ice rich worlds and rocky planets with envelopes is responsible for the radius valley (e.g. \citealt{2018MNRAS.479L..81R, 2020A&A...643L...1V, 2022ApJ...939L..19I, 2024NatAs...8..463B}), but the debate is far from conclusive at this stage (e.g. \citealt{2025ApJ...991L..16W}). Determining the detailed composition of these small exoplanets could thus help to understand the origin of this planetary population.

One way to probe the composition of material that formed exo-planets directly is through the atmospheres of white dwarfs (e.g. \citealt{2024A&A...691A.352W, 2025MNRAS.542..293R}). White dwarfs show only hydrogen/helium absorption lines, but some times, when exo-planetary materials (e.g exo-comets, exo-planetesimals, exo-planets) fall onto the white dwarfs, their composition can be inferred, as it alters the various absorption lines of the white dwarf. Shortly after the event, the (planetary) material sinks downwards and only the pure H/He absorption lines are visible again. Mostly, the composition of this material is consistent with planetesimals in the solar system \citep{2019Sci...366..356D}.

On the other hand, condensation calculations using different stellar abundances as input to constrain the composition of potential rocky planets show a large variety in the composition of these rocky planets (e.g. \citealt{2010Icar..205..321B, 2022A&A...660A..85J, 2025A&A...703A.218S}), which become even larger when planetary interactions are included (e.g. \citealt{2025A&A...694A.303H}). These different stellar abundances not only influence the composition of rocky planets, but may also dramatically alter the water content of sub-Neptunes due to large variations in the stellar C/O ratio \citep{2020A&A...633A..10B, 2023arXiv230105034C}.

An alternative way to extract the detailed composition of exoplanetary material  is related to stellar abundance measurements. \citet{2009ApJ...704L..66M} suggested that the formation of the terrestrial planets can take away more refractory (with high condensation temperature $T_{\rm cond}$) compared to volatile elements (with low $T_{\rm cond}$) during their formation process, leading to a lower refractory abundance in atmospheres of stars compared to the volatile elements. This hypothesis could also explain the abundances of the Sun, which shows a deficiency in refractory elements compared to the majority of solar twins (e.g. \citealt{2009ApJ...704L..66M, 2010A&A...521A..33R, 2018ApJ...865...68B, 2026arXiv260701699B}), which presumably did not form inner terrestrial planets.

Furthermore, wide binary or co-moving stars (collectively defined as 'co-natal stars' in this work) should form from the same parent molecular cloud and should thus contain the same elemental composition. However, studies found small abundance differences containing also a trend with condensation temperature between wide binary stars (e.g. \citealt{2014ApJ...790L..25T, 2014MNRAS.442L..51L, 2015A&A...582A..17S, 2016ApJ...819...19T}), indicating that the pattern found by \citet{2009ApJ...704L..66M} could be common. On a statistical level, a small fraction of these co-natal stars show differences in their abundance pattern \citep{2021NatAs...5.1163S, 2024Natur.627..501L}.

The abundance differences can not be explained with atomic diffusion (e.g. \citealt{2017ApJ...840...99D}), which would result in an abundance difference that scales with the atomic weight, whereas the observations show different trends with refractory and volatile elements (see table~\ref{tab:stellarabundances}) that can not be explained by atomic diffusion alone (e.g. \citealt{2024Natur.627..501L}). Instead the abundance differences can be attributed to the ingestion of planetary material into the stellar convective zone (e.g. \citealt{2021NatAs...5.1163S, 2024Natur.627..501L}).

The ingested mass that caused the measured stellar abundance differences was mostly compared to the composition of the Earth \citep{2001E&PSL.185...49A}, as a first benchmark. However, it is well known that stars with different stellar abundances form planets with different compositions (e.g. \citealt{2021Sci...374..330A, 2021Eleme..17..235P, 2024RvMG...90..259G}). We thus expand on this approach by using the observed detailed stellar abundances of the individual components of co-natal star systems (derived in \citealt{2023MNRAS.526.2181Y}) to calculate the composition of planets that may be ingested by these stars to explain the relative difference in their measured abundances.

In this work, our main assumption is that the stellar abundance differences between stars of a co-natal pair can be explained by the accretion of planet(s) by one (or both) of the stellar components.
Planet engulfment during the gas-disc phase would also add material to the stellar envelope, however at these stages stars are still mostly convective, so that unrealistically large amounts of material would need to be accreted (e.g. \citealt{2018A&A...618A.132K, 2018MNRAS.479.3690B, 2020A&A...641A..73H, 2020MNRAS.493.5079B}). Accretion of material directly from the protoplanetary discs during its evolution, suffers the same limitation: that the material is accreted early and thus only has minimal (if at all) influence on the present-day stellar abundances (e.g. \citealt{2023A&A...676A..87H}).

N-body planet formation simulations of close-in planetary systems show that systems are initially born in a resonant configuration, which are then broken by instabilities over time scales of a few 10 to 100 Myrs (e.g. \citealt{2017MNRAS.470.1750I, 2019arXiv190208772I, 2023A&A...674A.178B}) during which planets could be ingested by their central star. This picture is confirmed by observations, which indicate that initially nearly all inner planetary systems are in a resonance configuration, where consequent instabilities break these chains over time scales of several 10-100Myrs \citep{2024AJ....168..239D}. We thus make the assumption that observable planet engulfment happens after at least 50 Myr of system evolution, when the stellar convective zone has settled to values of 1-2.5\% of the stellar mass (e.g. \citealt{2011ApJS..192....3P, 2020A&A...641A..73H, 2023ApJS..265...15J}), allowing planetary engulfment to explain the stellar abundance differences. Furthermore, as the composition of planets varies from rocky super-Earth to potential volatile rich sub-Neptunes (e.g. \citealt{2019PNAS..116.9723Z, 2021JGRE..12606639B, 2025ApJ...991L..16W, 2026arXiv260415891N}), planetary engulfment has the potential to naturally explain different levels of stellar enrichment of volatile (e.g. C, O) and refractory (e.g. Fe, Si, Mg) elements.

Here we adopt the sample of co-natal stars from \citet{2024Natur.627..501L}, together with our chemical model, to determine the mass and composition of planets that need to be ingested by the stars to explain the measured stellar abundances differences. This will allow us to determine the mass and composition of planets that were formed around stars with different abundances from an observational perspective. Our aims are to (i) understand how massive the planets have to be that are engulfed into their central star and (ii) to retrieve the composition of the engulfed planets.

Our paper is structured as follows. We first discuss our methods (section~\ref{sec:methods}). We then show the results of our full sample in section~\ref{sec:full} before we discuss the implications of our simulations on the detailed planetary compositions in section~\ref{sec:composition}. We then discuss our results in a more general context (section~\ref{sec:discussion}) before we conclude in section~\ref{sec:conclusion}.

\section{Methods}
\label{sec:methods}

In this section, we describe the details of our chemical partitioning model and how we calculate the changes of the stellar abundances via the addition of planetary material to the stellar convective zone. We close this section with a description of our stellar sample and a description of how we use the stellar sample to determine the masses and compositions of the planets that are accreted into the stellar convective zone.

\subsection{Chemical partitioning model}

\begin{table*}
\centering
\caption{Condensation temperatures and volume mixing ratios of the chemical species.}
\begin{tabular}{c c c}
\hline\hline
Species (Y) & $T_{\text{cond}}$ {[}K{]} & $v_{\text{Y, with C}}$  \\ \hline
CO & 20  & 0.2 $\times$ C/H   \\
CH$_4$ & 30 & 0.1 $\times$ C/H   \\
CO$_2$ & 70 & 0.1 $\times$ C/H  \\
H$_2$S & 150 & 0.1 $\times$ S/H   \\
H$_2$O & 150 & O/H - (3 $\times$ MgSiO$_3$/H + 4 $\times$ Mg$_2$SiO$_4$/H + CO/H + 2 $\times$ CO$_2$/H + 3 $\times$ Fe$_2$O$_3$/H)  \\
C (carbon grains) & 631 & 0.6 $\times$ C/H\\
FeS & 704 & 0.9 $\times$ S/H   \\
Mg$_2$SiO$_4$ & 1354 & Mg/H - Si/H \\
Fe$_2$O$_3$ & 1357 & 0.5 $\times$ (Fe/H - 0.9 $\times$ S/H)  \\
MgSiO$_3$ & 1500 & Mg/H - 2$\times$(Mg/H - Si/H) \\
\hline
\end{tabular}
\tablefoot{Condensation temperatures for molecules are taken from \citet{2003ApJ...591.1220L}. For Fe$_2$O$_3$ the condensation temperature for pure iron is adopted \citep{2003ApJ...591.1220L}. Volume mixing ratios $v_{\rm Y}$ (i.e. by number) are adopted for the species as a function of disk elemental abundances. We expand here on the different mixing ratios from \citet{2020A&A...633A..10B}. If the temperature at which the planet forms is above the condensation temperature of FeS, we attribute all Fe into Fe$_2$O$_3$ and the sulfur into the gaseous H$_2$S.}
\label{tab:comp}
\end{table*}

We list here the solar abundances (table~\ref{tab:solar_values}) taken from \citet{2009ARA&A..47..481A}, which we use as standard input to derive the observed stellar abundances \citep{2023MNRAS.526.2181Y} as well as for our planet engulfment model. These abundances can be scaled to the measured stellar abundances using the [X/H] notation.

We now construct a chemical partitioning model (see table~\ref{tab:comp}) that includes the different elements listed in table~\ref{tab:solar_values}. Our model is based on \citet{2020A&A...633A..10B}, where a similar approach was taken in \citet{2015A&A...580L..13S}. In this chemical partitioning model, we distribute the atoms into solids/ices, because protoplanetary discs do not contain atoms directly like stars. We first distribute carbon, sulfur and the refractory elements (Mg, Si, Fe) to their corresponding solids and then use the remaining oxygen to form water. For example, 20\% of the available carbon forms CO, 10\% CH$_4$, 10\% CO$_2$ and the remaining 60\% of carbon are in refractory carbon grains (see table~\ref{tab:comp}).

Given our model, the formation temperature of the planet sets its composition. If the planet were to form at temperatures below 20K, it would contain all elements in proportions that correspond to the solar abundances, because all elements are in solid form. If the planet forms at a higher temperature, the composition will change, as some of the ices are not available in solid form any more and can thus not be incorporated into the forming planet. For instance, a planet forming at a temperature above 150K will contain only small amounts of oxygen (that are stored in Mg$_2$SiO$_4$, Fe$_2$O$_3$, and MgSiO$_3$), while a planet forming below 150K will also contain H$_2$O, and thus much more oxygen (see also \citealt{2020A&A...633A..10B}). This changes the mass fractions of the different elements within the planet, see appendix~\ref{ap:abundance} for more details.

If the disc temperature at which the planet accretes material is larger than the condensation temperature of FeS (704K), the planet will not accrete sulfur any more, because FeS is destroyed into refractory iron grains and H$_2$S vapor (e.g. \citealt{2023A&A...676A..52T}), which is then not accreted by the planet, because we only follow solid accretion in our model. If the disc temperature is larger than the evaporation temperature of FeS, we attribute the remaining Fe into Fe$_2$O$_3$. A planet fully forming at these temperatures (above 704 K) will thus contain no sulfur, but the fully available iron content.

\subsection{Stellar abundances}

The change of the stellar abundances is determined by the material of the planet that is dissolved in the stellar convective zone. The change of the abundances of the stellar convective zone by a given element is then determined by the amount of material that already is contained in the convective zone and how much is added by the engulfed planet. This abundance difference due to planet engulfment, $\Delta [X/H]$, is calculated in the following way
\begin{equation}
\label{eq:engulf}
 \Delta [X/H] = \log_{\rm 10} \left( \frac{M_{\rm X, CZ} + M_{\rm X, P}}{M_{\rm tot, CZ}} \times \frac{\mu_{\rm sun}}{m_{\rm X}} \frac{1}{X/H} \right) \ .
\end{equation}
Here $M_{\rm X, CZ}$ and $M_{\rm X, P}$ denote the mass of the probed element $X$ (corresponding to Fe, Mg, Si, S, C, or O) in the convective zone and within the planet that is accreted into the convective zone. $M_{\rm tot, CZ}$ marks the total mass of the convective zone. $\mu_{\rm sun}$ corresponds to the mean molecular weight of the sun (equal to 1.3455) and $m_{\rm X}$ corresponds to the molecular weight of element $X$, where $X/H$ denotes the original elemental abundances (e.g. for the sun listed in tab.~\ref{tab:solar_values}, but of course modified for the individual stars, see tab.~\ref{tab:stellarabundances}). We use here a value of 1 or 2.5\% of the stellar mass for the total mass of the stellar convective zone, corresponding to the values used in \citet{2024Natur.627..501L}, which were determined from stellar evolution models. The corresponding masses of the convective-zone are consistent with stellar evolution models. We note that the adpoted values of $M_{\rm CZ}$ primarily affects the inferred total mass of the engulfed material, rather than its relative chemical composition, which is the main focus of our study. We therefore do not expect more detailed estimates of $M_{\rm CZ}$ to alter our main conclusions. We note that eq.~\ref{eq:engulf} is an adaption of a similar method used in \citet{2018MNRAS.479.3690B} and \citet{2023A&A...676A..87H}, who studied the impact of the accretion of solid and gaseous protoplanetary disc material onto stellar abundances rather than the ingestion of planets.

The abundance in the stellar convective zone thus changes when material is accreted. If the planetary composition contains all elements  (Fe, Mg, Si, S, C and O) in a solar ratio, then the stellar abundance difference will change equally for all elements once the planet is engulfed. This happens if the planet formed at a temperature below 20K, because all molecules and thus elements are frozen out and thus are fully incorporated into the planet. At higher formation temperatures, less molecules are frozen out and can thus not be accreted as solids into the planet any more. Therefore the stellar abundances differences between the different elements of different volatility will change once the planet is engulfed. For example, if the planet formed at a formation temperature of 720K, the planet will not contain any carbon and any sulfur, because both, carbon and sulfur, are not available in solid form any more at these temperatures (see table~\ref{tab:comp}). The corresponding stellar abundances difference of carbon and sulfur will thus remain at zero, whereas abundance differences for the two stars for O, Mg, Si, and Fe will exist, as these elements are incorporated into the engulfed planet.

We note that \citet{2024Natur.627..501L} determined measurements of many more elements compared to what we use here in our analysis. However, the trend of these elements scales with their volatility. All refractory elements show generally the same trend, which is why we limit ourselves to Fe, Mg, and Si. As the volatile elements (O, C) and also sulfur show very different trends, we focus on these elements in our analysis. In addition, these elements constitute $\sim$ 80\% of the total heavy element mass of the sun, where another $\sim$15\% originates from N and Ne, which do not play a role in planet formation as they are extremely volatile and only included in gaseous planets and are additionally not tracked in \citet{2024Natur.627..501L}. The mass and composition determinations of our inferred planets thus harbor and error of at most a few percent.

\subsection{Stellar sample}

The stellar sample originates from \cite{2024Natur.627..501L}. In particular, we focus here on the 7 systems already identified to have potentially engulfed a planet. \cite{2024Natur.627..501L} identified planet-engulfment candidates through Bayesian model comparison between a planet-ingestion model, an atomic-diffusion model, and a flat/null model. Systems were classified as strong engulfment candidates when the planet-ingestion model was significantly favoured, based on the adopted Bayesian evidence threshold. The abundance difference $\Delta [X/H]$ is defined as the abundance differences of star B minus star A:
\[
 \Delta [X/H] = [X/H]_{\rm B} - [X/H]_{\rm A}
\]
A negative abundance difference implies that Star A has a higher elemental abundances compared to star B, implying that Star A ingested a planet in our model. In order to calculate the composition of the engulfed planet, we start with the hypothesis that the star with the overall lower Fe abundance corresponds to the original composition of the molecular cloud that formed the co-natal pair, stars A and B. We then proceed to calculate how much planetary material needs to be added to give the composition of the co-natal companion that is enriched compared to the first star. This allows us to calculate the plausible composition of the planet(s) that are swallowed by their host star. We note that in principle both stars could have ingested a planet, influencing the abundance pattern of the reference star in the process. However, even if this would be the case, the relative abundance difference still needs to be explained by additional material that is accreted onto one of the stars.

\subsection{Modelling of the stellar abundance differences}

Put together, we aim to understand the stellar abundance differences within co-natal stellar pairs by adding material of planetary origin into one of the stars of the co-natal pair to match the measured abundance differences. Under the assumption that both stars in the co-natal pair formed from the same parent molecular cloud (and thus with the same stellar abundances), we use the star with the overall lower abundances as a starting point. We note that the study by \cite{2024A&A...682L..23S} indicated that the abundances differences in some of the (giant) binary systems are caused by primordial inhomogeneities rather than engulfment events. Nevertheless, the sample selection of \citet{2024Natur.627..501L} ensures that the selected stars are co-natal and should thus share the identical initial chemical compositions, in contrast to field stars or solar twins.

We further make the assumption that the star with the higher abundances has swallowed a planet, which is then consequently dissolved in its convective envelope and increased the initial stellar abundance to the currently observed abundances (starting from the lower abundances). We therefore use the stellar abundance measurements of the star with the lower abundances to calculate a stoichomentric model (see table~\ref{tab:comp}) that will give the elemental composition of a planet formed at a given temperature (see appendix~\ref{ap:abundance}) and is then ingested by the star.

In order to match the observed stellar abundance differences, we probe how different planetary masses that are added to the star with the lower stellar abundances impact the stellar abundance differences via eq.~\ref{eq:engulf}. Furthermore, we additionally vary the formation temperature of the planet, as this regulates the composition of the planet and with this the abundance differences in the convective zones of the star once the planet is ingested. Our aim is two-fold: We want to (i) understand how massive the planets have to be that are accreted into the stellar convective zone\footnote{The planetary masses have been determined using an Earth like composition in \citet{2024Natur.627..501L} and we thus expect only small variations in respect to the planetary masses compared to their work.} and (ii) the composition of the planet that is accreted, because the composition sets the individual elemental abundance differences of the stars. Once we match the stellar abundance differences we can retrieve the planetary mass and composition. We note that equation~\ref{eq:engulf} is applied to all 6 elements investigated in our study simultaneously, allowing us to disentangle how much of the individual elements is accreted onto the star to explain the abundance differences. The summation of these elements gives the total planetary mass. The composition indicates the formation location of the planets (e.g. a planet forming interior to the water ice line will contain relatively less oxygen compared to a planet of the same mass that formed beyond the water ice line, see Fig.~\ref{fig:Massfraction}), giving insights into the general formation of planetary systems.

Close-in exoplanets of the sub-Neptune class harbor a small (1-2\% of their total mass) envelope, which can also contain heavy elements and not only hydrogen and helium (e.g. \citealt{2024arXiv240303325B, 2025A&A...701A.184K, 2025arXiv250712622H}). Due to its small mass, we do not take the envelope contribution into account, because the errors in the relative stellar abundance determination are generally of the order of $\approx 5$\% and above, indicating that the influence of the planetary envelope on the stellar abundances is minimal.

\section{Origin of the stellar abundance differences}
\label{sec:full}

We discuss in this section how planet engulfment can explain the abundance differences of the co-natal systems of our sample (table~\ref{tab:stellarabundances}). We will first start with the simplest scenario that one star of the co-natal pair engulfed one planet that formed completely between different evaporation fronts, thus harboring discrete compositions. In case this scenario does not allow a match to the observed abundance differences, we expand the scenario to include the migration of planets across different evaporation fronts during their formation and to a multi-planet scenario, where both components of the co-natal stellar pair ingest a planet. We note that the masses of the planets that need to be engulfed by the central star depend also on the mass of the convective zone (eq.~\ref{eq:engulf}), which is calculated via stellar models according to the individual stellar parameters (e.g. mass, $T_{\rm eff}$, Z, age), see \citealt{2024Natur.627..501L}. This results in different convective envelope masses for each co-natal pair. We explain this procedure in detail for the co-natal pair 69 (see \citealt{2024Natur.627..501L}), and then follow with the rest of the sample. We chose pair 69, because it features abundance measurements for all our elements of interest (O, C, S, Mg, Si, Fe) and provides a clear example of the different assumptions regarding mass and origin of the planets that will also be made for the other pairs.

\subsection{Benchmark case (pair 69)}

Pair 69 features abundances measurements for all our elements of interest (C, O, S, Fe, Mg, Si), see table~\ref{tab:stellarabundances}. The stellar convective zone corresponds to 1\% of the stellar mass for pair 69.

\subsubsection{Single planet model}

We investigate three different types of planets formed at 3 different locations in the disc, where different formation locations result in different planetary compositions. We investigate planets forming at 100K, 300K, and 720K. These correspond to planets forming exterior to the water ice line, resulting in very water and thus oxygen rich planets (100K), planets forming just interior to the water ice line, where planets will contain less oxygen (300K), as well as planets forming in the very inner disc, where FeS and Carbon grains have evaporated, resulting in a planet that does not contain any carbon and sulfur.

\begin{figure*}
 \centering
 \includegraphics[scale=0.6]{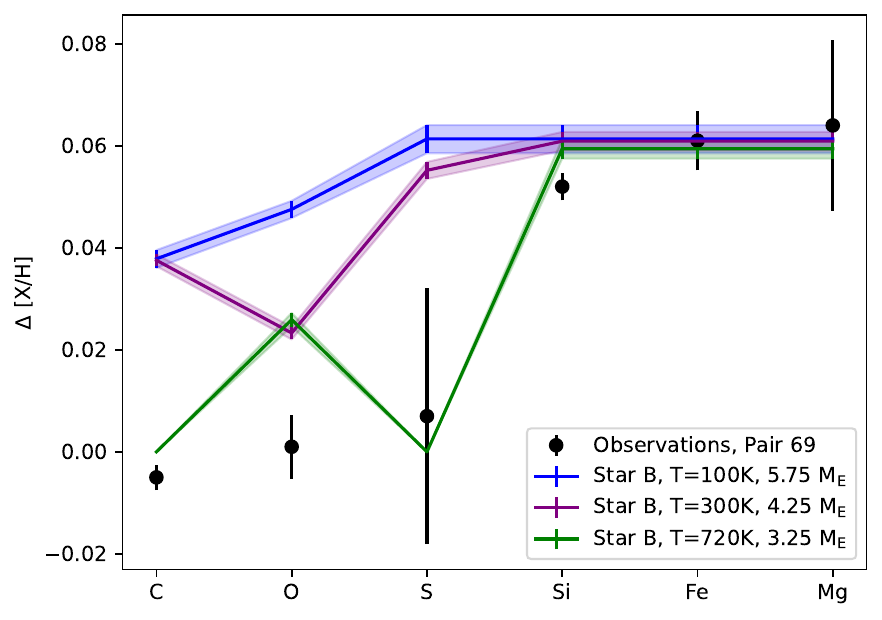} \quad
 \includegraphics[scale=0.6]{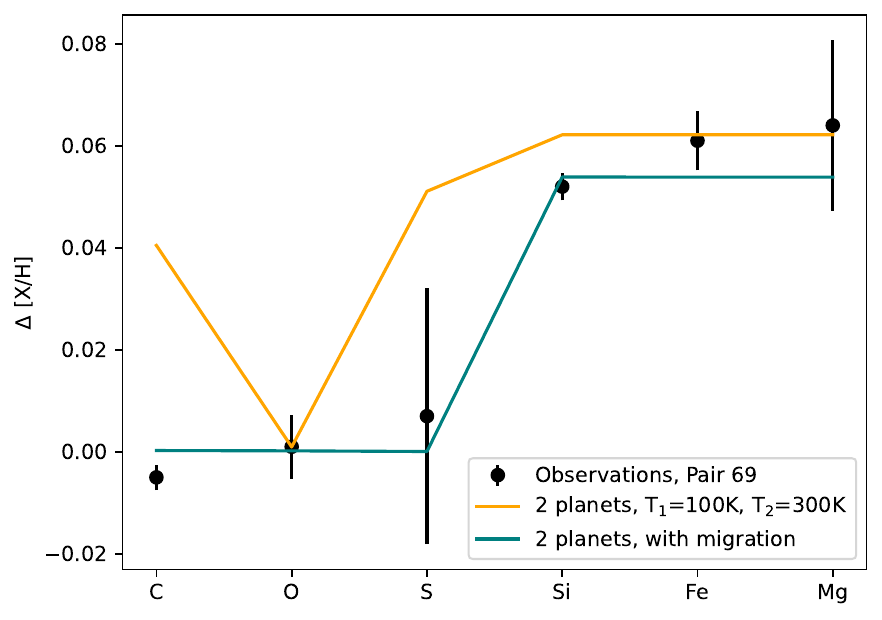}
 \caption{Left: Abundance differences of stellar pair 69 (dots), as well as the calculated change of the stellar abundance of star B if planets of different masses and different formation temperatures (and thus compositions) have been accreted by the star. The error bars of the curves correspond to the errors in planetary composition caused by the errors in the abundance measurements of the star B (see table~\ref{tab:stellarabundances}). The mass of the convective zone is 1\% of the stellar mass. Right: Same, but now two planets are used to explain the measured abundance differences. One planet originates from beyond the water ice line and one from interior to the water ice line. Both planets are then engulfed by their host star. The plot shows a scenario where both planets do not migrate and another scenario where the planets can migrate.
   \label{fig:pair69lowMZ}
   }
\end{figure*}

The exact stellar abundance differences caused in a stellar atmosphere will depend not only on the composition of the planet, but also on the mass of the planet (eq.~\ref{eq:engulf}), as well as on the size of the convective zone (see Appendix~\ref{ap:convection}). We thus adjust the mass of the engulfed planet to match the abundance difference of the refractory elements (Si, Fe, Mg) between Star A and Star B, because these elements are determined with a higher robustness (due to a larger availability of lines) compared to the volatiles\footnote{Additionally, some other co-natal pairs miss abundance measurements of oxygen and/or sulfur.}. In addition, volatile elements are distributed in many different molecules within protoplanetary discs (e.g. table~\ref{tab:comp}), adding another layer of complexity for volatile elements. We show in Fig.~\ref{fig:pair69lowMZ} the results of these simulations.

By construction, the abundance difference of the refractory elements is matched within the error bars. While the planets just forming exterior and interior of the water ice line fail to match the observed carbon, oxygen, and sulfur abundance differences, the planet forming in the very inner disc regions matches pretty well the carbon and sulfur abundance differences (which are close to zero), but would result in a too large oxygen abundance difference. This is caused by the fact that the engulfed planet contains oxygen within its rock forming elements (see table~\ref{tab:comp}).

While this simple model fails to explain the oxygen abundance difference between the two stars, we can identify a few key components of the planetary ingestion model in respect to the stellar abundance differences of oxygen versus the refractory components.

\begin{itemize}
 \item The change of the planetary composition and consequently of the stellar abundances due to planet engulfment that are caused by the errors of the determination of the individual stellar abundances (table~\ref{tab:stellarabundances}) have a smaller effect on the abundance differences than the formation location of the planets. This indicates that high-precision error determinations of stellar abundances are necessary to support the planet engulfment scenario.

 \item If the abundance difference of the refractory elements is a factor of $\approx$1.3 larger than the oxygen abundance difference, the engulfed planet would need to form exterior to the water ice line and contain a large fraction of oxygen\footnote{In order to have a zero abundance difference between oxygen and the refractory components, the planet would need to form exterior to the CO ice line, resulting in the accretion of all solids containing oxygen}.

 \item If the abundance difference of the refractory components compared to oxygen is about a factor of $\approx 2-2.5$ larger, the ingested planet formed interior to the water ice line, where only roughly $\approx 40\%$ of the oxygen is incorporated into the planet, due to the limited amount of oxygen within the refractory solids (e.g. in Fe$_2$O$_3$).

 \item A much larger difference between the refractory and oxygen abundance differences can not be explained with a model that contains a single non-migrating planet, because the planet will always contain oxygen due to the refractory solids (e.g. Fe$_2$O$_3$), resulting in a minimal oxygen to refractory ratio (see above).
\end{itemize}

As pair 69 shows a close to zero abundance difference in oxygen, while simultaneously a large refractory element abundance difference, it can not be explained with a single planet model, even if planetary migration is included. This is caused by the fact that planets, independently, of how much they migrate, will always contain oxygen, preventing to generate an abundances difference of oxygen that is closer to zero. We will now investigate if a scenario where both stars engulf a planet could help to explain the observed abundance pattern.

\subsubsection{Engulfment of two planets}

We remain with the assumption that the initial stellar abundance corresponds to star B, even though both stars in this proposed scenario will engulf a planet. This, in principle, would alter the stellar abundances of both stars and we would not have a reference point for the individual stellar abundances available, as for the situation where only one star would ingest a planet. We think this assumption is justified, because we are interested in the relative abundance difference between the stars and not in the absolute stellar abundances.

We show in Fig.~\ref{fig:pair69lowMZ} the stellar abundance differences if two planets are ingested by their respective host stars. In the first case, the planets do not migrate (namely, they stay within the respective evaporation fronts). We first aim to explain the close to zero oxygen abundance difference between the stars. We thus make the assumption that the two planets formed interior and exterior to the water ice line, so that they contain different amounts of oxygen (see Appendix~\ref{ap:abundance}) and thus also different amounts of the refractory elements. In our example, the stars swallowed a planet with 6 Earth masses from beyond the water ice line and the other star a 9.5 Earth masses from interior of the water ice line. These masses have been calibrated to achieve the negligible oxygen abundance difference between the two stars, while at the same time giving a matching refractory abundance difference.

However, as shown in Fig.~\ref{fig:pair69lowMZ}, this scenario results in a too large abundance difference between the two stars in respect to carbon and sulfur. On the other hand, we have seen in our model (Fig.~\ref{fig:pair69lowMZ}) that planets that form in the very inner disc regions will contain no carbon and no sulfur. We thus expand our model to include the migration of one of the planets towards the inner disc to match in addition to the oxygen abundance difference, the carbon and sulfur abundance difference.

In the migration case the planet from beyond the water ice line is still kept at 6 Earth masses (for simplicity). The other planet first needs to accretes 4.5 Earth masses of material from interior to the water ice line and exterior to the C-evaporation front. As the 4.5 Earth masses are accreted from interior to the water ice line, the planet will contain less oxygen, but relatively more carbon compared to the 6 Earth mass planet formed exterior to the water ice line. The masses are now calibrated in such a way that if the stars were to swallow these respective planets that the carbon abundance difference is zero.

The smaller 4.5 Earth mass planet then migrates inwards and accretes 0.375 Earth masses from interior of the C-evaporation front, but exterior to the FeS evaporation front. This results in an zero abundance differences for sulfur. The planet then further migrates inwards and accretes 3.0 Earth masses of material from the very inner disc region, which contains no sulfur and no carbon, but still oxygen and refractory elements. Consequently, this also results in a zero abundance difference of oxygen between the two stars if they were to swallow these specific planets. By invoking the migration scenario for one of the planets to contain material from different disc regions to calibrate the carbon, sulfur and oxygen abundances, we automatically (without fine tuning) match the refractory abundance differences of Si, Fe, and Mg. We note that it is not guaranteed that we would match the refractory abundance differences, because for the migration model, we have chosen one given planet and have then calibrated the mass of the planet around the other star by matching the volatile components, which - in principle - could result in a mismatch of the refractory component. Matching the refractory component thus gives additional support for the migration scenario of at least one of the planets.

We conclude that the dichotomy between the volatile and refractory element abundance differences between the stars can be explained by the two planet scenario (incl. migration) for pair 69.

\subsection{Full sample}
\label{subsec:full}

\begin{figure*}
 \centering
 \includegraphics[scale=0.375]{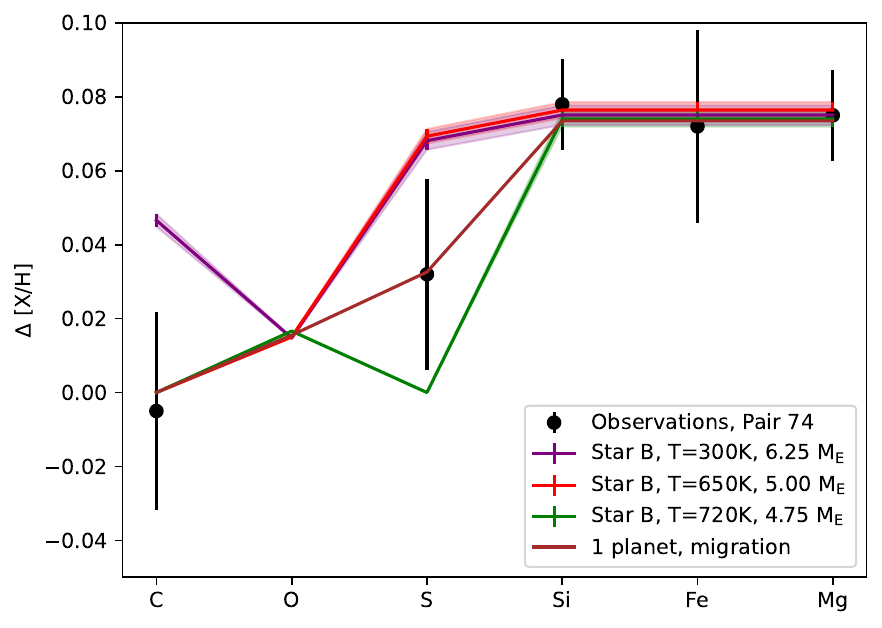} \quad
 \includegraphics[scale=0.375]{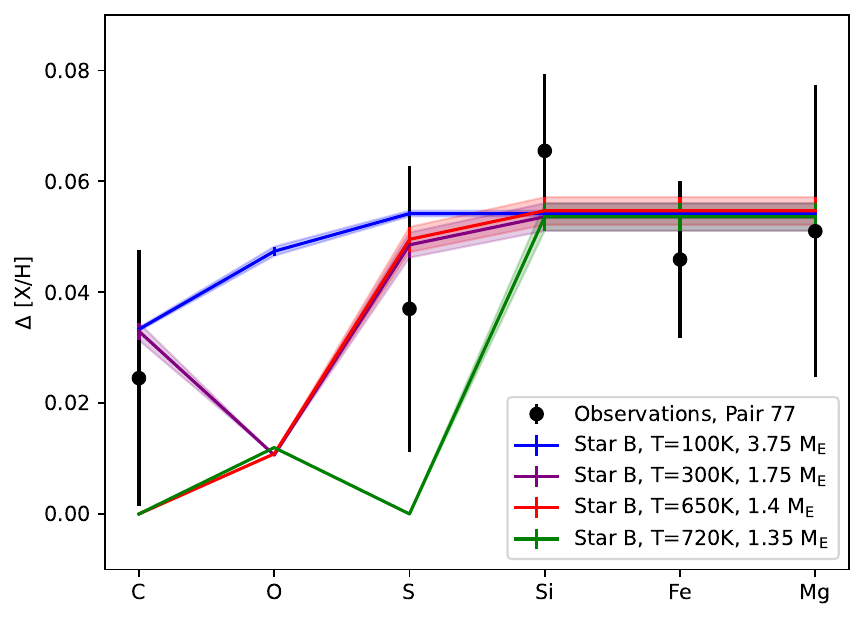} \quad
 \includegraphics[scale=0.375]{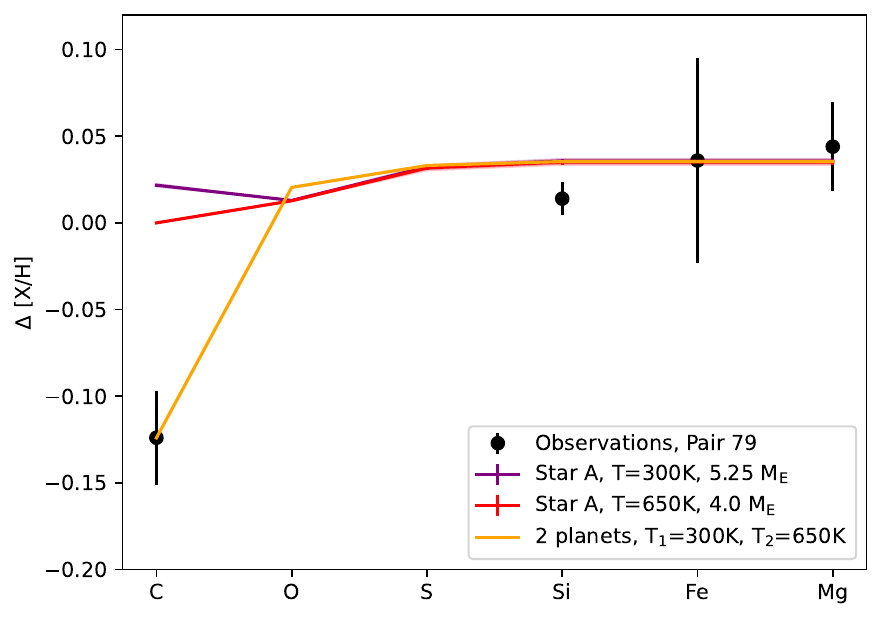} \quad
 \includegraphics[scale=0.375]{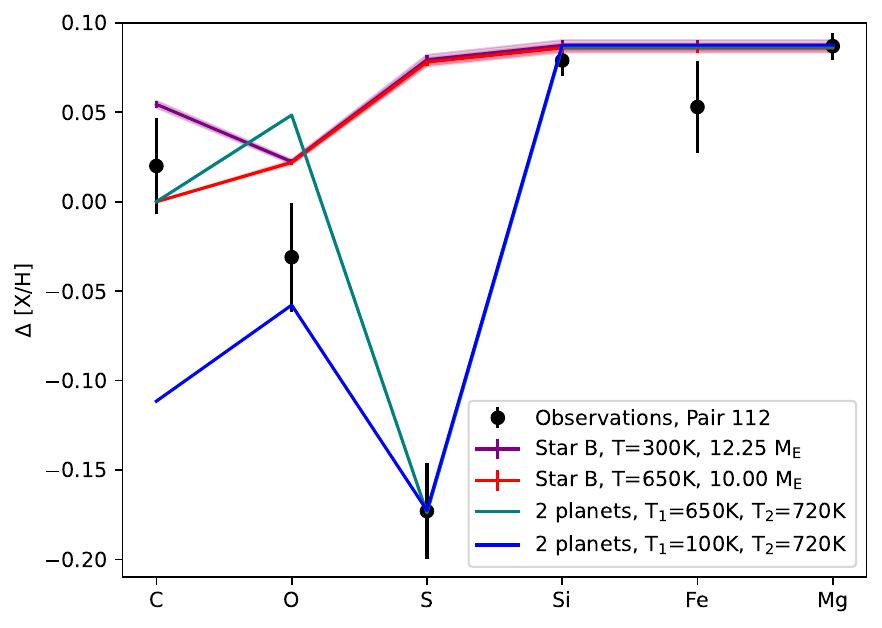} \quad
 \includegraphics[scale=0.375]{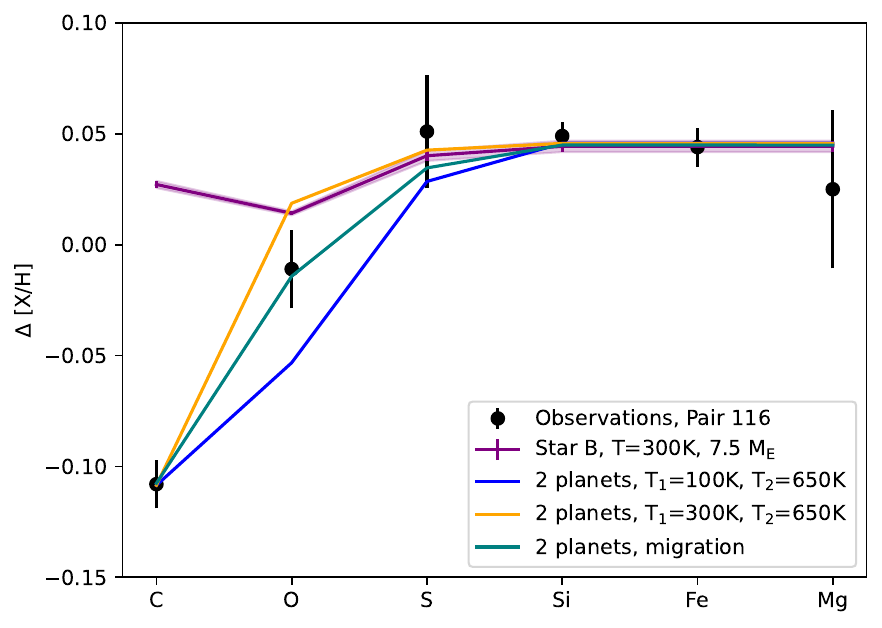} \quad
 \includegraphics[scale=0.375]{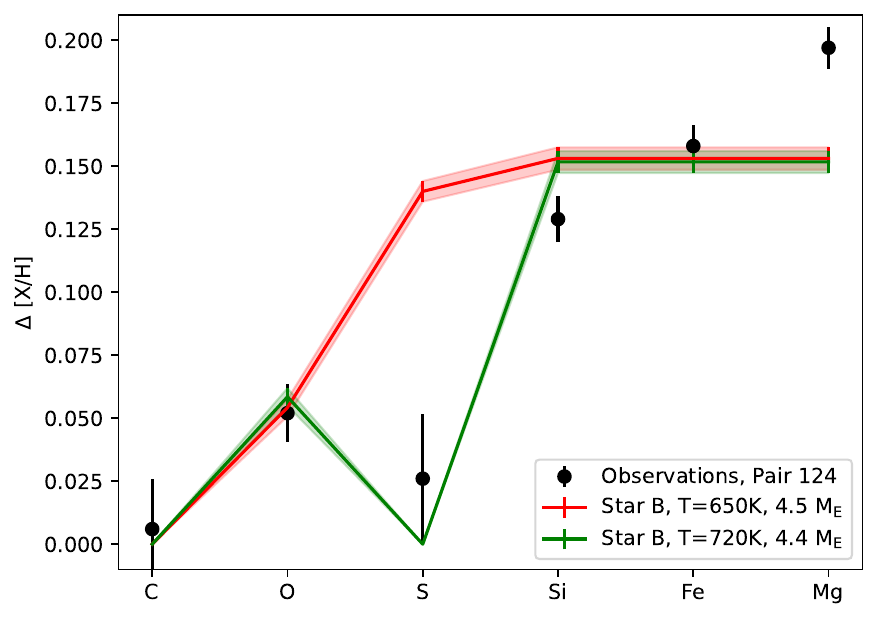}
 \caption{Stellar abundance differences of pair 74, 77, 79, 112, 116, and 124 (top left to bottom right) as well as simple models to explain the abundance differences, either with single non migrating planets formed at different disc locations (and thus temperatures, where different colors inidicate different formation temperatures of the planets), single migrating planet or multiple planets. Some of the pairs do not show a measured oxygen and/or sulfur abundance. Our planetary ingestion model can not explain the abundances differences of pair 79 and pair 112 (middle row).
   \label{fig:pairs}
   }
\end{figure*}

We show in Fig.~\ref{fig:pairs} the remaining 6 stellar pairs and their corresponding abundance differences, as well as our best fit planetary engulfment models. As explained above, we test different planetary formation scenarios, where the planets form in different regions (and thus at different temperatures) in the protoplanetary disc, which allows the planets to harbor different compositions, which then set the stellar abundance differences. Two systems require single planets formed at specific temperatures (pairs 77 and 124), while one system requires a single migrating planet (pair 74) and one additional system requires 2 migrating planets (pair 124). For two systems (pairs 79 and 112) our analysis is inconclusive, because the measure stellar abundances could not be matched completely. We list the general outcome of our simulations in table~\ref{tab:systemsexplain}.

\begin{table*}
\centering
\caption{Summary of our preferred formation channels for the different systems.}
\small{
\begin{tabular}{ccccc}
Pair  & preferred formation channel & origin of planets & Planet mass & $M_{\rm CZ} [M_\star]$ \\ \hline
69 & 2 planets, migration & 1 exterior, 1 interior to H$_2$O line & 6.0 M$_{\rm E}$, 7.8 M$_{\rm E}$ & 0.01 \\
74 & 1 planet, migration & interior to C line; migrate across FeS line & 4.75 M$_{\rm E}$ & 0.025 \\
77 & 1 planet, no large-scale migration & interior to H$_2$O line & 1.75 M$_{\rm E}$ & 0.01 \\
79  & not conclusive & - & - & 0.025 \\
112 & not conclusive & - & - & 0.025 \\
116 & 2 planets, migration & 1 exterior, 1 interior to H$_2$O line & 30 M$_{\rm E}$, 25 M$_{\rm E}$ & 0.025 \\
124 & 1 planet, no large-scale migration & interior to FeS line & 4.4 M$_{\rm E}$ & 0.01 \\
  \hline
\end{tabular}}
\tablefoot{\normalsize We list the mass of the convective zone $M_{\rm CZ}$ (see \citealt{2024Natur.627..501L}) in the last column. We note that the mass of the convective zone influences the mass of the planet that is ingested, but does not influence the composition and thus formation location of the planets (see Appendix~\ref{ap:convection}).}
\label{tab:systemsexplain}
\end{table*}

Pair 74 has no measured oxygen abundance difference, limiting somewhat the analysis. The models with single planets forming at specific temperatures fail to explain the measured abundance differences, as they either contain too much or too little sulfur. This is caused by the fact that sulfur is tied to iron via FeS (table~\ref{tab:comp}). Instead, if the planet form first beyond the FeS evaporation front and then migrates across it, allows a match to the measured abundance differences. Constraining the oxygen abundance difference of this stellar pair would give additional constraints to our model.

Pair 77 also has no measured oxygen abundance difference. Using single planets that form at specific temperatures allow to match the observed abundance differences for this stellar pair. In particular, we find that the planet that was ingested by Star A of the system has to form interior to the water ice line, as this allows a slightly better match to the sulfur abundance measurement, compared to the planet forming exterior to the water ice line. Again, constraining the oxygen abundance difference would help to discriminated between these two options.

Pair 79 features no measured oxygen and sulfur abundance difference, limiting the analysis. The pair features a negative carbon abundance difference with a positive Si, Fe, and Mg abundance difference, indicating that a solution where both stars ingest a planet is needed. However, this model can not directly match the observed abundance differences, because the abundance differences of Si, Mg, and Fe are different, but our model only allows for a combined accretion of all three components, which are refractory in nature (see table~\ref{tab:comp}).

Pair 112 features a negative sulfur and oxygen abundance difference paired with positive abundance differences of C, Mg, Si, and Fe. A one planet model could explain the abundances differences of C, Mg, Si, and Fe, but fails to explain the O and S abundance differences. The 2 two planet models fail either with the oxygen or carbon abundance differences, while matching the other constraints. A model that fits the C, O and S abundance differences (not displayed) fails to match the Si, Fe, and Mg abundance differences. This is caused by the fact that sulfur is mostly tied to the accretion of Fe, indicating that both elemental abundance difference should be shifted in the same way by planetary ingestion. We can thus not safely conclude if planetary ingestion is at play here.

Pair 116 features a negative carbon and oxygen abundance difference paired with a positive abundance difference of Si, Fe, and Mg. This can not be explained by a model that only allows one start to ingest a planet. Instead a two planet model, where one planet migrates across the water ice line can explain the measured abundance differences.

Pair 124 features a close to zero abundance difference of carbon and sulfur, while the other elemental abundance differences have a positive value. The measured abundance differences can thus be explained by a model where one of the stars ingests a planet that formed interior to the FeS evaporation front and thus does not contain any carbon or sulfur. We note that in principle also a planet that formed exterior to the FeS evaporation front and then migrated across it could explain the abundance differences, but we opt with the simpler model that does not require planet migration. Thus, pair 124 is the only system where the measured abundances differences can be explained by a single non migrating planet.

\section{Planetary compositions}
\label{sec:composition}

\begin{figure*}
 \centering
 \includegraphics[scale=0.5]{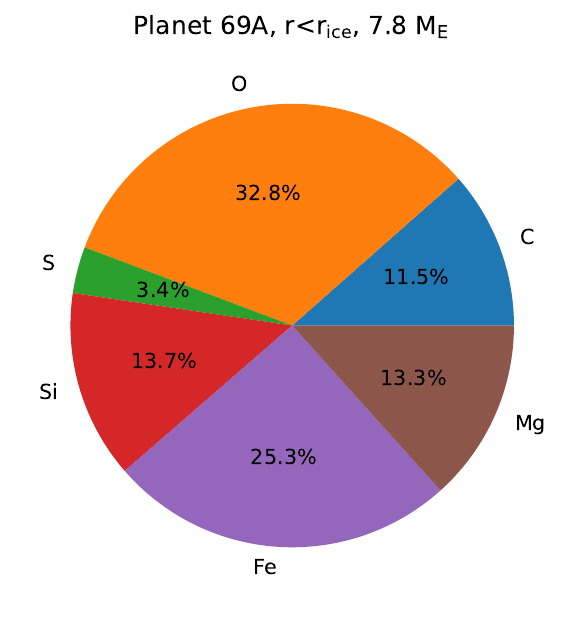} \quad
 \includegraphics[scale=0.5]{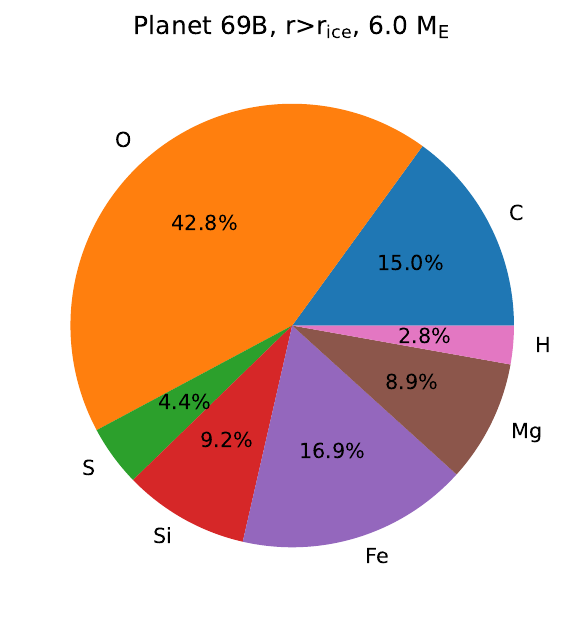} \quad
 \includegraphics[scale=0.5]{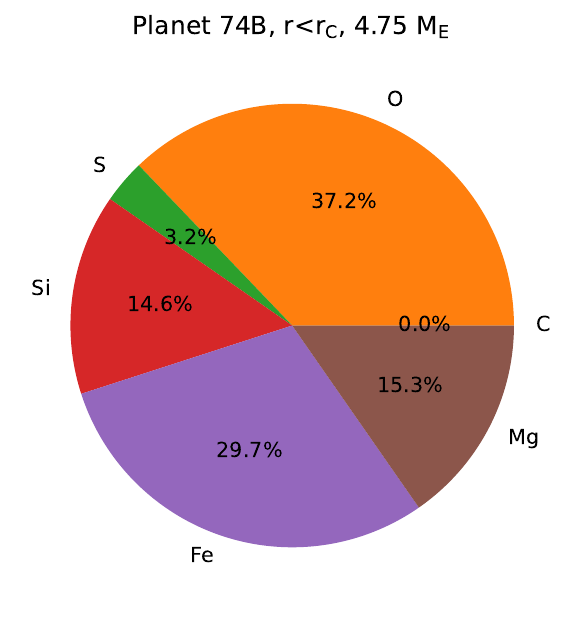} \quad
 \includegraphics[scale=0.5]{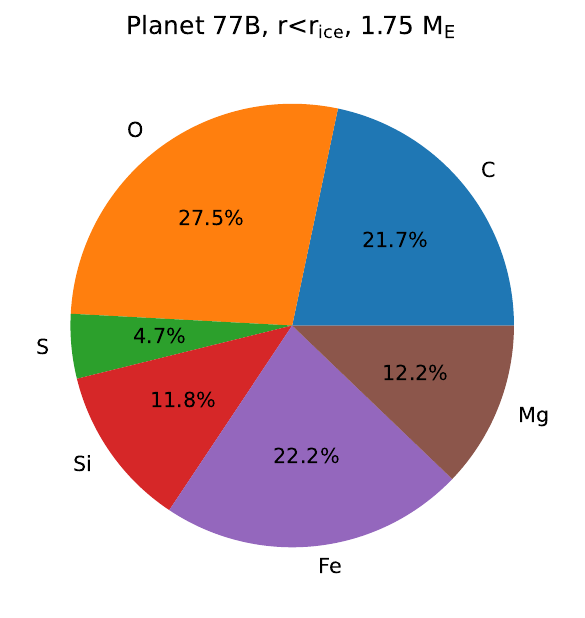} \quad
 \includegraphics[scale=0.5]{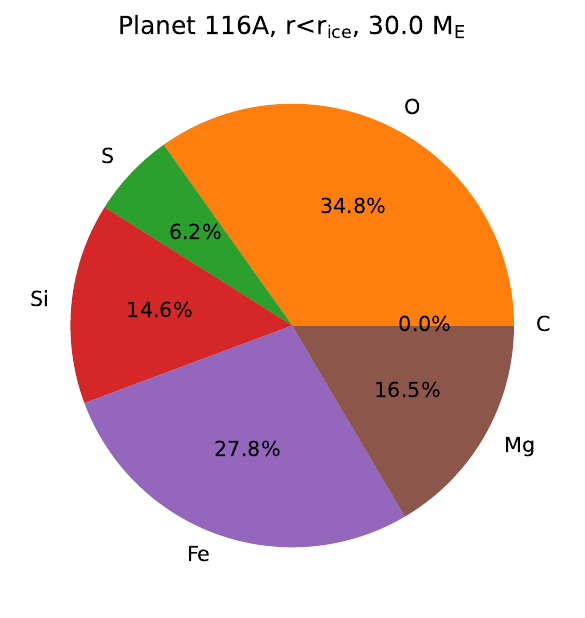} \quad
 \includegraphics[scale=0.5]{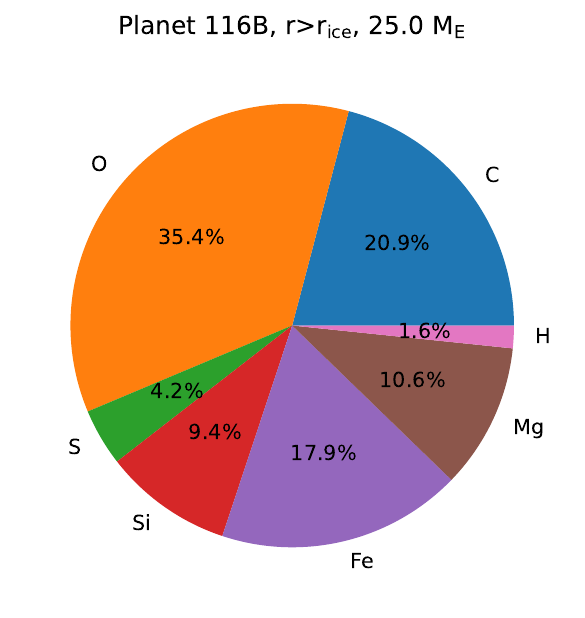} \quad
 \includegraphics[scale=0.5]{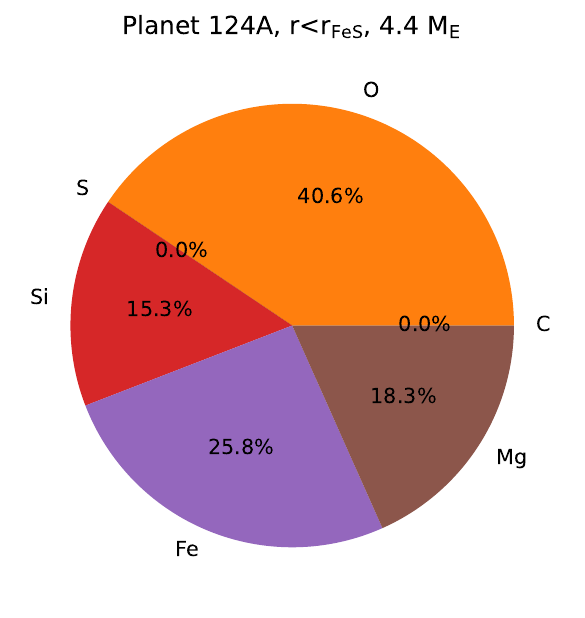} \quad
 \caption{Mass fractions of the planets that are ingested by the different stars. The label A and B behind the planetary pairs name, indicate that both stars need to ingest a planet. We do not show the composition of the potential planets for pair 79 and 112, as our analysis was inconclusive (Fig.~\ref{fig:pairs} and table~\ref{tab:systemsexplain}). We only show here the composition based on the stellar abundance differences without the error bars, because the errors do not significantly alter the planetary composition (see Appendix~\ref{ap:abundance}).
   \label{fig:Massfractionpairs}
   }
\end{figure*}

In Fig.~\ref{fig:Massfractionpairs} we show the composition of the planets that are ingested by the stars within our model. Some systems (pair 69 and 116) require that both stars ingest a planet (marked with A and B), while we do not show the potential planetary compositions for pairs 79 and 112, as our analysis was inconclusive in respect to the planetary composition (see table~\ref{tab:systemsexplain}). The differences in planetary masses are caused not only by the formation location of the planets, but also by the stellar convective zone, where a larger stellar convective zone requires a larger planetary mass to explain the observed stellar abundance differences.

The composition of the potential planets that are ingested by their central star depend on three effects: (i) the stellar abundances of the host star, which regulates the composition of the disc, (ii) the formation location of the planets, which regulate their composition, and (iii) the stellar abundance differences which regulate how much planet forming material is needed and if material from across different evaporation fronts is necessary. Due to the diversity of stellar compositions and stellar abundance differences (table~\ref{tab:stellarabundances}), our simulations show a large variety of planetary compositions (Fig.~\ref{fig:Massfractionpairs}).

It is generally expected that planets that form beyond the water ice line should harbor larger fractions of oxygen compared to planets forming interior to the water ice line (e.g. \citealt{2020A&A...633A..10B, 2023arXiv230105034C}). Our analysis shows that this is not necessarily always the case. The planet 116B forms beyond the water ice line, but its oxygen mass fraction is similar to planets that formed interior to the water ice line around other stars (e.g. Planet 69A, Planet 74). This is caused by the fact that the reference star in pair 116 is largely enriched in carbon compared to oxygen, facilitating the formation of CO and CO$_2$ compared to H$_2$O, resulting in a lower overall water abundance within the protoplanetary disc in which the planets form (see also \citealt{2020A&A...633A..10B, 2023arXiv230105034C}) and thus consequently also in less oxygen within the forming planets. Our analysis shows oxygen mass fractions from around 27 to 43\% for all potential planets.

Similar effects can be seen for the iron mass fraction. Even though the planet around pair 124 forms in the very inner disc regions (interior to the FeS evaporation front), it contains less iron compared to the planet 116A, which forms exterior to the FeS evaporation front and should contain less iron (as part of the planet is made of sulfur, reducing the overall mass fractions of all other elements). This effect is again caused by the differences in stellar abundances, which influences the iron content of the planets (see also \citealt{2021Sci...374..330A}). The iron mass fractions are generally lower for planets that form exterior to the water ice line, as these planets contain (on average) larger amounts of oxygen, compared to planets forming interior to the water ice line. Our analysis shows iron mass fractions ranging from 17 to 30\%, which is lower than the iron fractions predicted in \cite{2021Sci...374..330A}. This is caused by the fact that our model includes iron in its oxidized form (Fe$_2$O$_3$), allowing a significant accretion of oxygen into the planet compared to the model used in \cite{2021Sci...374..330A}.

Generally our analysis shows a wide range of planetary compositions, showing the importance of taking stellar abundances into account when calculating planetary compositions (e.g. ~\citealt{2021Sci...374..330A, 2022A&A...660A..85J, 2025A&A...703A.218S}). The study by \cite{2025A&A...693A..64A} investigated the abundances differences of a binary stellar pair consisting of a main-sequence star and a white dwarf, which was ingesting material. \cite{2025A&A...693A..64A} could determine the differences of the refractory material accreted onto the white dwarf compared to the abundances of the main-sequences F-star. They found that the material ingested by the white dwarf was depleted in moderate refractories (Mg, Si, and Fe) relative to refractory materials (Ca, Ti), indicating that not all refractory materials are incorporated evenly into forming planetesimals (or planets, see \citealt{2026arXiv260824407W}), opening up another (complementary) pathway to determine the detailed composition of planet forming material in special systems.

\section{Discussion}
\label{sec:discussion}

\subsection{Limitations of our model}

Our model is based on the assumption that the measured abundance differences between stars in co-natal pairs are caused by the ingestion of planets into one or both components of the pair. The idea of planetary ingestion is based on the fact that atomic diffusion can not explain the observed abundance differences (e.g. \citealt{2024Natur.627..501L}). In contrast, the study by \citet{2024A&A...682L..23S} seems to indicate that the differences in the stellar abundances could be caused by primordial inhomogeneities rather than engulfment of planetary material. Their study shows only positive abundance differences for all elements (except Al, Li, and Ba). However, O, Mg, and Si are all $\alpha$-elements and produced in the same stellar environments (e.g. \citealt{1957RvMP...29..547B}), indicating that the abundance differences between two stars should all scale in the same direction (positive or negative). To explain the measured abundance differences from \citet{2024Natur.627..501L}, where oxygen can have a negative abundances difference, while Mg and Si have a positive abundance difference between the two stars, indicates that another effect is at play at least in these systems. Planetary ingestion could indeed provide this.

We focus here only on six elements, whereas the full co-natal stellar sample \citep{2024Natur.627..501L} contains more elements. However, these mostly refractory elements show generally the same abundance differences as Fe, Mg, and Si. At the same time, this implies that our chemical model is limited, as more exotic elements like CaS or MgS (e.g. \citealt{2026arXiv260315955Z}) could exist and harbor large fractions of sulfur (relevant for e.g. pair 112, Fig.~\ref{fig:pairs}). We also make the assumption that e.g. iron is fully oxidized (e.g. in the form of Fe$_2$O$_3$) compared to other models (e.g. \citealt{2023A&A...673A..17M}), where iron is not oxidized, leading to higher iron mass fractions and lower oxygen mass fractions within the planets for these models.

In our work we use different masses of the convective zone (either 1 or 2.5\% of the stellar mass, based on the calculations in \citealt{2024Natur.627..501L}), whereas in reality the mass of the convective zone also changes in time (e.g. \citealt{2018A&A...618A.132K, 2020A&A...641A..73H}). While these estimates of the convective zone are subject to error and depend also on the overall metallicity of the stars themselves, they do not affect the general trend of our conclusions. If stars with a larger convective zone had been modelled with smaller convective zones, they would need to ingest lower mass planets, but the formation region of these planets in the disc would not be affected, because this is entirely based on the individual elemental abundance differences and not by the size of the convective zone, see Appendix~\ref{ap:convection}.

Our study is limited to the idea of the accretion of (rocky) small mass planets, whereas also higher mass planets, dominated by gas, could be ingested by stars. Gaseous planets (incl. sub-Neptunes), on the other hand, might be enhanced in volatile elements (C, O, N) compared to refractories in their atmospheres (e.g. \citealt{2021A&A...654A..71S, 2021ApJ...909...40T, 2023A&A...679L...7D, 2025NatAs...9..845E, 2026A&A...706A..30O}). While carbon and oxygen are also relatively abundant in solid components with relatively high evaporation temperatures, nitrogen is mostly present in the disc as N$_2$ and NH$_3$, with very low evaporation temperatures. It would thus predominately incorporated into gaseous planets (e.g. Jupiter is enhanced by a factor of 3-4 times solar in nitrogen, \cite{2016arXiv160604510A}). Thus an observed nitrogen abundance difference between stars in co-natal pairs could hint at the ingestion of gaseous planets. If the planetary atmosphere would be eroded during the ingestion process and blown away, no nitrogen abundances difference would be observable. However, the differences in the stellar abundances with respect to the refractory elements, as well as for carbon and oxygen would remain.

\subsection{Implications for planet formation}

In this work, we attributed the abundance differences of co-natal stellar pairs to the engulfment of planets above a few Earth masses, which can be either rocky in nature or contain larger fraction of volatiles (corresponding to sub-Neptunes). The planetary mass that needs to be added to the central star is larger, if the stellar convective zone is larger. As discussed above, gravitational interactions between chains of inner super-Earth planets can result in instabilities (e.g. \citealt{2017MNRAS.470.1750I, 2019arXiv190208772I}), where planets can also be thrown into the central star. The composition of these planets then drives the stellar abundance difference pattern. While determining the exact composition of close-in super-Earths and mini-Neptunes via transit and RV measurements is degenerate (e.g. \citealt{2015A&A...577A..83D, 2019PNAS..116.9723Z}), our analysis opens up a pathway for more accurate composition determinations similar to the accretion of material onto white dwarfs (e.g. \citealt{2024A&A...691A.352W, 2025MNRAS.542..293R}).

We do not find any trends with stellar mass, as the stellar masses are all roughly solar in our sample (tab.~\ref{tab:stellarparameters}). We also do not recover an obvious trend with the stellar metallicities, which cover a wide range ([Fe/H] values ranging from -0.3 to +0.2). However, the stellar metallicities are consistent with the occurrence of super-Earths and mini-Neptunes (e.g. \citealt{2014Natur.509..593B}). Considering that roughly 30-50\% of solar like stars should host close-in super-Earths and/or mini-Neptunes (e.g. \citealt{2013ApJ...766...81F, 2018AJ....156...24M}, our conclusion that a majority of the stars in our sample (corresponding to $\approx$7\% of all stars in the \cite{2024Natur.627..501L} sample) ingested a planet is reasonable.

Statistically speaking, the planetary radius valley at around 1.8 $R_{\rm E}$ divides the small planets into predominately rocky super-Earths and outer sub-Neptunes that could be rocky with hydrogen/helium envelopes or water rich planets (e.g. \citealt{2024arXiv240303325B, 2025A&A...701A.184K}). Planet formation simulations speculate that the sub-Neptune population should contain a fraction of water worlds (e.g. \citealt{2022ApJ...939L..19I, 2024NatAs...8..463B}). Even though our analysis is limited by the small sample size (only 4 of our 7 stellar pairs contain oxygen abundance measurements), we find that a minimum of 2 stars should have ingested a planet originating from beyond the water ice line (see table~\ref{tab:systemsexplain}). This is in agreement with the before mentioned planet formation simulations and thus supports a large scale migration of planets.

As planets do not fall onto stars by themselves, they require outer perturbers that will drive the inner planets into their central star. These perturbers could originate within the close-in chains of super-Earths themselves or exterior to the water ice line (e.g. \citealt{2019arXiv190208772I, 2023A&A...674A.178B, 2024A&A...692A.246B}). Thus, the remains of the planetary systems after the scattering events that drove the abundance differences of the stars, should be observable as either close-in super-Earths/sub-Neptunes or outer ice/gas giants.

\section{Conclusion}
\label{sec:conclusion}

The differences of the measured stellar abundances in co-natal pairs are in agreement with the planet engulfment hypothesis, because it can explain the abundance differences of the stars with the planetary composition derived from the individual stellar abundances. An exception is pair 112, where the difference in the refractory elements (negative sulfur), but positive Si, Mg, and Fe paired with a negative oxygen abundance makes it difficult to fit a two component/planet solution (see table~\ref{tab:systemsexplain}). We were also unable to achieve a definite conclusion for pair 79.

While the sample is small (only 7 pairs), it allows already to identify several different formation pathways:
\begin{itemize}
 \item No large-scale migration: Some abundance difference patterns can be explained easily by the ingestion of a single planet around one of the components of the co-natal pair, where the planet probably formed in the inner disc region, namely interior to the water ice line (e.g. pair 77, pair 124). This scenario is characterized by about a factor of 3 larger refractory abundance difference compared to the oxygen or carbon abundance differences. As the planet forms from solid material in the inner disc regions, it only accretes a small fraction of carbon and oxygen, as large fractions of these materials have already been evaporated into the gas phase (see table~\ref{tab:comp}). Consequently, the star ingesting this planet will be enriched in refractory elements by about a factor of 3 more than in oxygen compared to its stellar companion that did not engulf a planet.

 \item Migration: Pair 74 (Fig.~\ref{fig:pairs}) shows about a factor of two difference of the sulfur abundance compared to the Si, Mg and Fe abundances. If the planet formed entirely exterior to the FeS evaporation line, it would accrete sulfur in a similar manner as Fe, Mg, and Si and the stars would thus show a similar abundance difference pattern for all these elements (like pair 116). If the planet formed entirely interior to the FeS evaporation line, it would contain no sulfur. Thus, in order to explain the factor two difference between the sulfur abundance differences and the refractory element difference, the planet would need to accrete part of its mass exterior to the FeS evaporation line and part of it interior to it, implying that the planet migrated through the disc. Alternatively, of course, the same star could have swallowed two (or more) planets, which formed either completely interior or exterior to the FeS evaporation line. In any case, different formation locations of the planets are necessary to explain the abundance differences.

 \item Multiple planet ingestion: Pair 69, 79, 112, and 116 show zero or negative abundance differences for the volatile elements (oxygen and carbon), but positive abundance differences for the refractory elements. As the ingestion of one planet onto one of the stars can only cause a completely positive abundance difference in Fe, Mg, Si and O\footnote{Oxygen is the key element here, because a planet could be completely devoid of carbon and sulfur in our model, if it formed interior to the C-grain and FeS evaporation line. However, that planet will still contain oxygen.}, a two planet model is needed. The idea of the two planet model is that one planet formed exterior to the water ice evaporation line and accreted a large fraction of oxygen (see Appendix~\ref{ap:abundance}), but less refractory components, while the planet around the other star formed interior the water ice line and thus accreted less oxygen relative to the refractory elements. If both stars accrete these planets, we can achieve a negative oxygen abundance and simultaneously a positive refractory element abundance. Similarly, we can explain a negative carbon abundance if one planet formed interior to the carbon-grain evaporation front, while the other planet formed exterior to it.

\end{itemize}

Even though our study is limited in the number of systems, due to the fact that not many systems with detailed abundances differences are determined \citep{2021NatAs...5.1163S, 2024Natur.627..501L}, our model suggests several pathways to explain the observed relative abundance differences of stars in co-natal pairs. These different pathways reflect the complexity of planet formation and show that the remnants of the planetary instabilities that drive the re-arrangement of super Earth systems (e.g. \citealt{2019arXiv190208772I, 2023A&A...674A.178B}) can be observed. Furthermore, it shows that super-Earth planets are required to form at different formation locations (close-in, interior and exterior to the water ice line), implying that migration is a necessary ingredient for the formation of planetary systems.

\begin{acknowledgements}
 FL is supported by the National Natural Science Foundation of China (NSFC; grant No. 2025000209). HSW is supported by the Carlsberg Foundation's Semper Ardens grant (‘FIRSTATMO’).
\end{acknowledgements}

\bibliographystyle{aa}
\bibliography{Stellar}

@article{2003ApJ...591.1220L,
	author = "Lodders, K.",
	journal = "ApJ",
	pages = "pp. 1220--1247",
	title = "{Solar system abundances and condensation temperatures of the elements}",
	volume = "591",
	year = "2003"
}

@article{2013ApJ...766...81F,
	author = "Fressin, F. and Torres, G. and Charbonneau, D. and Bryson, S.~T. and Christiansen, J. and Dressing, C.~D. and Jenkins, J.~M. and Walkowicz, L.~M. and Batalha, N.~M.",
	journal = "ApJ",
	pages = "id.81",
	title = "{The False Positive Rate of Kepler and the Occurrence of Planets}",
	volume = "766",
	year = "2013"
}

@article{2014Natur.509..593B,
	author = "Buchhave, L.~A. and Bizzarro, M. and Latham, D.~W. and Sasselov, D.~D. and Cochran, W.~D. and Endl, M. and Isaacson, H. and Juncher, D. and Marcy, G.W.",
	journal = "Nature",
	pages = "pp.593--595",
	title = "{Three regimes of extrasolar planet radius inferred from host star metallicities}",
	volume = "509",
	year = "2014"
}

@article{2015A&A...580L..13S,
	author = "Santos, N.~C. and Adibekyan, V. and Mordasini, C. and Benz, W. and Delgado-Mena, E. and Dorn, C. and Buchhave, L.~A. and Mortier, A. and Pepe, F. and Santerne, A. and Sousa, S. G. and Udry, S.",
	journal = "A\&A",
	pages = "id.L13",
	title = "{Constraining planet structure from stellar chemistry: the cases of CoRoT-7, Kepler-10, and Kepler-93}",
	volume = "580",
	year = "2015"
}

@article{1995Natur.378..355M,
	author = "Mayor, M. and Queloz, D.",
	journal = "Nature",
	pages = "pp.355--359",
	title = "{A Jupiter-mass companion to a solar-type star}",
	volume = "378",
	year = "1995"
}

@article{2017MNRAS.470.1750I,
	author = "Izidoro, A. and Ogihara, M. and Raymond, S.~N. and Morbidelli, A. and Pierens, A. and Bitsch, B. and Cossou, C. and Hersant, F.",
	journal = "MNRAS",
	pages = "pp. 1750--1770",
	title = "{Breaking the chains: hot super-Earth systems from migration and disruption of compact resonant chains}",
	volume = "470",
	year = "2017"
}

@article{2014ApJ...790L..25T,
	author = "{Tucci Maia}, M. and Melendez, J. and Ramirez, I.",
	journal = "ApJL",
	pages = "id. L25",
	title = "{High Precision Abundances in the 16 Cyg Binary System: A Signature of the Rocky Core in the Giant Planet}",
	volume = "790",
	year = "2014"
}

@article{2014MNRAS.442L..51L,
	author = "Liu, F. and Asplund, M. and Ramirez, I. and Melendez, J.",
	journal = "MNRAS letters",
	pages = "p.L51--L55",
	title = "{A high-precision chemical abundance analysis of the HAT-P-1 stellar binary: constraints on planet formation}",
	volume = "442",
	year = "2014"
}

@article{2016ApJ...819...19T,
	author = "Teske, J. K. and Khanal, S. and Ramirez, I.",
	journal = "ApJ",
	pages = "id. 19",
	title = "{The Curious Case of Elemental Abundance Differences in the Dual Hot Jupiter Hosts WASP-94A and B}",
	volume = "819",
	year = "2016"
}

@article{2009ARA&A..47..481A,
	author = "Asplund, M. and Grevesse, N. and Sauval, A. J. and Scott, P.",
	journal = "ARAA",
	pages = "pp.481--522",
	title = "{The Chemical Composition of the Sun}",
	volume = "47",
	year = "2009"
}

@article{2015A&A...582A..17S,
	author = "Saffe, C. and Flores, M. and Buccino, A.",
	journal = "A\&A",
	pages = "id. A17",
	title = "{HD 80606: searching for the chemical signature of planet formation}",
	volume = "582",
	year = "2015"
}

@article{2009ApJ...704L..66M,
	author = "Melendez, J. and Asplund, M. and Gustafson, B.~A. and Yong, D.",
	journal = "ApJL",
	pages = "L66--L70",
	title = "{The Peculiar Solar Composition and Its Possible Relation to Planet Formation}",
	volume = "704",
	year = "2009"
}

@article{2017AJ....154..109F,
	adsurl = "http://adsabs.harvard.edu/abs/2017AJ....154..109F",
	archiveprefix = "arXiv",
	author = "{Fulton}, B.~J. and {Petigura}, E.~A. and {Howard}, A.~W. and {Isaacson}, H. and {Marcy}, G.~W. and {Cargile}, P.~A. and {Hebb}, L. and {Weiss}, L.~M. and {Johnson}, J.~A. and {Morton}, T.~D. and {Sinukoff}, E. and {Crossfield}, I.~J.~M. and {Hirsch}, L.~A.",
	doi = "10.3847/1538-3881",
	eid = "109",
	eprint = "1703.10375",
	journal = "\aj",
	month = sep,
	pages = "109",
	primaryclass = "astro-ph.EP",
	title = "{The California-Kepler Survey. III. A Gap in the Radius Distribution of Small Planets}",
	volume = 154,
	year = 2017
}

@article{2013ApJ...775..105O,
	adsurl = "http://adsabs.harvard.edu/abs/2013ApJ...775..105O",
	archiveprefix = "arXiv",
	author = "{Owen}, J.~E. and {Wu}, Y.",
	doi = "10.1088/0004-637X",
	eid = "105",
	eprint = "1303.3899",
	journal = "\apj",
	month = oct,
	pages = "105",
	primaryclass = "astro-ph.EP",
	title = "{Kepler Planets: A Tale of Evaporation}",
	volume = 775,
	year = 2013
}

@article{2017ApJ...847...29O,
	adsurl = "http://adsabs.harvard.edu/abs/2017ApJ...847...29O",
	archiveprefix = "arXiv",
	author = "{Owen}, J.~E. and {Wu}, Y.",
	doi = "10.3847/1538-4357",
	eid = "29",
	eprint = "1705.10810",
	journal = "\apj",
	month = sep,
	pages = "29",
	primaryclass = "astro-ph.EP",
	title = "{The Evaporation Valley in the Kepler Planets}",
	volume = 847,
	year = 2017
}

@article{2018AJ....156...24M,
	adsurl = "http://adsabs.harvard.edu/abs/2018AJ....156...24M",
	archiveprefix = "arXiv",
	author = "{Mulders}, G.~D. and {Pascucci}, I. and {Apai}, D. and {Ciesla}, F.~J.",
	doi = "10.3847/1538-3881/aac5ea",
	eid = "24",
	eprint = "1805.08211",
	journal = "\aj",
	month = jul,
	pages = "24",
	primaryclass = "astro-ph.EP",
	title = "{The Exoplanet Population Observation Simulator. I. The Inner Edges of Planetary Systems}",
	volume = 156,
	year = 2018
}

@article{2019arXiv190208772I,
	adsurl = "https://ui.adsabs.harvard.edu/abs/2021A\&A...650A.152I",
	archiveprefix = "arXiv",
	author = "{Izidoro}, Andr{\'e} and {Bitsch}, Bertram and {Raymond}, Sean N. and {Johansen}, Anders and {Morbidelli}, Alessandro and {Lambrechts}, Michiel and {Jacobson}, Seth A.",
	doi = "10.1051/0004-6361/201935336",
	eid = "A152",
	eprint = "1902.08772",
	journal = "\aap",
	month = jun,
	pages = "A152",
	primaryclass = "astro-ph.EP",
	title = "{Formation of planetary systems by pebble accretion and migration. Hot super-Earth systems from breaking compact resonant chains}",
	volume = "650",
	year = 2021
}

@article{2020A&A...633A..10B,
	adsurl = "https://ui.adsabs.harvard.edu/abs/2020A\&A...633A..10B",
	archiveprefix = "arXiv",
	author = "{Bitsch}, Bertram and {Battistini}, Chiara",
	doi = "10.1051/0004-6361/201936463",
	eid = "A10",
	eprint = "1911.09725",
	journal = "\aap",
	month = jan,
	pages = "A10",
	primaryclass = "astro-ph.EP",
	title = "{Influence of sub- and super-solar metallicities on the composition of solid planetary building blocks}",
	volume = "633",
	year = 2020
}

@article{2019PNAS..116.9723Z,
	adsurl = "https://ui.adsabs.harvard.edu/abs/2019PNAS..116.9723Z",
	archiveprefix = "arXiv",
	author = "{Zeng}, Li and {Jacobsen}, Stein B. and {Sasselov}, Dimitar D. and {Petaev}, Michail I. and {Vanderburg}, Andrew and {Lopez-Morales}, Mercedes and {Perez-Mercader}, Juan and {Mattsson}, Thomas R. and {Li}, Gongjie and {Heising}, Matthew Z. and {Bonomo}, Aldo S. and {Damasso}, Mario and {Berger}, Travis A. and {Cao}, Hao and {Levi}, Amit and {Wordsworth}, Robin D.",
	doi = "10.1073/pnas.1812905116",
	eprint = "1906.04253",
	journal = "Proceedings of the National Academy of Science",
	month = may,
	number = "20",
	pages = "9723--9728",
	primaryclass = "astro-ph.EP",
	title = "{Growth model interpretation of planet size distribution}",
	volume = "116",
	year = 2019
}

@article{2019MNRAS.484..712D,
	adsurl = "https://ui.adsabs.harvard.edu/abs/2019MNRAS.484..712D",
	archiveprefix = "arXiv",
	author = "{Dorn}, C. and {Harrison}, J.~H.~D. and {Bonsor}, A. and {Hands}, T.~O.",
	doi = "10.1093/mnras/sty3435",
	eprint = "1812.07222",
	journal = "\mnras",
	month = mar,
	number = "1",
	pages = "712--727",
	primaryclass = "astro-ph.EP",
	title = "{A new class of Super-Earths formed from high-temperature condensates: HD219134 b, 55 Cnc e, WASP-47 e}",
	volume = "484",
	year = 2019
}

@article{2018MNRAS.479L..81R,
	adsurl = "https://ui.adsabs.harvard.edu/abs/2018MNRAS.479L..81R",
	archiveprefix = "arXiv",
	author = "{Raymond}, Sean N. and {Boulet}, Thibault and {Izidoro}, Andre and {Esteves}, Leandro and {Bitsch}, Bertram",
	doi = "10.1093/mnrasl/sly100",
	eprint = "1805.10345",
	journal = "\mnras",
	month = "Sep",
	number = "1",
	pages = "L81--L85",
	primaryclass = "astro-ph.EP",
	title = "{Migration-driven diversity of super-Earth compositions}",
	volume = "479",
	year = "2018"
}

@article{2020A&A...643L...1V,
	adsurl = "https://ui.adsabs.harvard.edu/abs/2020A\&A...643L...1V",
	archiveprefix = "arXiv",
	author = "{Venturini}, Julia and {Guilera}, Octavio M. and {Haldemann}, Jonas and {Ronco}, Mar\'{\i}a P. and {Mordasini}, Christoph",
	doi = "10.1051/0004-6361/202039141",
	eid = "L1",
	eprint = "2008.05513",
	journal = "\aap",
	month = nov,
	pages = "L1",
	primaryclass = "astro-ph.EP",
	title = "{The nature of the radius valley. Hints from formation and evolution models}",
	volume = "643",
	year = 2020
}

@article{2021A&A...654A..71S,
	adsurl = "https://ui.adsabs.harvard.edu/abs/2021A\&A...654A..71S",
	archiveprefix = "arXiv",
	author = "{Schneider}, Aaron David and {Bitsch}, Bertram",
	doi = "10.1051/0004-6361/202039640",
	eid = "A71",
	eprint = "2105.13267",
	journal = "\aap",
	month = oct,
	pages = "A71",
	primaryclass = "astro-ph.EP",
	title = "{How drifting and evaporating pebbles shape giant planets. I. Heavy element content and atmospheric C/O}",
	volume = "654",
	year = 2021
}

@article{2021ApJ...909...40T,
	adsurl = "https://ui.adsabs.harvard.edu/abs/2021ApJ...909...40T",
	archiveprefix = "arXiv",
	author = "{Turrini}, D. and {Schisano}, E. and {Fonte}, S. and {Molinari}, S. and {Politi}, R. and {Fedele}, D. and {Pani{\'c}}, O. and {Kama}, M. and {Changeat}, Q. and {Tinetti}, G.",
	doi = "10.3847/1538-4357/abd6e5",
	eid = "40",
	eprint = "2012.14315",
	journal = "\apj",
	month = mar,
	number = "1",
	pages = "40",
	primaryclass = "astro-ph.EP",
	title = "{Tracing the Formation History of Giant Planets in Protoplanetary Disks with Carbon, Oxygen, Nitrogen, and Sulfur}",
	volume = "909",
	year = 2021
}

@INCOLLECTION{2016arXiv160604510A,
       author = {{Atreya}, Sushil K. and {Crida}, Aurelien and {Guillot}, Tristan and {Lunine}, Jonathan I. and {Madhusudhan}, Nikku and {Mousis}, Olivier},
        title = "{The Origin and Evolution of Saturn, with Exoplanet Perspective}",
    booktitle = {Saturn in the 21st Century},
         year = 2018,
       editor = {{Baines}, Kevin H. and {Flasar}, F. Michael and {Krupp}, Norbert and {Stallard}, Tom},
        pages = {5-43},
          doi = {10.1017/9781316227220.002},
       adsurl = {https://ui.adsabs.harvard.edu/abs/2018stfc.book....5A}
}

@article{2021Sci...374..330A,
	adsurl = "https://ui.adsabs.harvard.edu/abs/2021Sci...374..330A",
	archiveprefix = "arXiv",
	author = "{Adibekyan}, Vardan and {Dorn}, Caroline and {Sousa}, S{\'e}rgio G. and {Santos}, Nuno C. and {Bitsch}, Bertram and {Israelian}, Garik and {Mordasini}, Christoph and {Barros}, Susana C.~C. and {Delgado Mena}, Elisa and {Demangeon}, Olivier D.~S. and {Faria}, Jo{\~a}o P. and {Figueira}, Pedro and {Hakobyan}, Artur A. and {Oshagh}, Mahmoudreza and {Soares}, B{\'a}rbara M.~T.~B. and {Kunitomo}, Masanobu and {Takeda}, Yoichi and {Jofr{\'e}}, Emiliano and {Petrucci}, Romina and {Martioli}, Eder",
	doi = "10.1126/science.abg8794",
	eprint = "2102.12444",
	journal = "Science",
	month = oct,
	number = "6565",
	pages = "330--332",
	primaryclass = "astro-ph.EP",
	title = "{A compositional link between rocky exoplanets and their host stars}",
	volume = "374",
	year = 2021
}

@article{1957RvMP...29..547B,
	adsurl = "https://ui.adsabs.harvard.edu/abs/1957RvMP...29..547B",
	author = "{Burbidge}, E. Margaret and {Burbidge}, G.~R. and {Fowler}, William A. and {Hoyle}, F.",
	doi = "10.1103/RevModPhys.29.547",
	journal = "Reviews of Modern Physics",
	month = "Jan",
	number = "4",
	pages = "547--650",
	title = "{Synthesis of the Elements in Stars}",
	volume = "29",
	year = "1957"
}

@article{2020A&A...641A..73H,
	adsurl = "https://ui.adsabs.harvard.edu/abs/2020A\&A...641A..73H",
	archiveprefix = "arXiv",
	author = "{Hoppe}, Richard and {Bergemann}, Maria and {Bitsch}, Bertram and {Serenelli}, Aldo",
	doi = "10.1051/0004-6361/201936932",
	eid = "A73",
	eprint = "2007.01297",
	journal = "\aap",
	month = sep,
	pages = "A73",
	primaryclass = "astro-ph.SR",
	title = "{The solar abundance problem and eMSTOs in clusters. Possible origins in the accretion of protoplanetary discs during stellar evolution}",
	volume = "641",
	year = 2020
}

@article{2023arXiv230105034C,
      author = {{Cabral}, N. and {Guilbert-Lepoutre}, A. and {Bitsch}, B. and {Lagarde}, N. and {Diakite}, S.},
        title = "{How the origin of stars in the Galaxy impacts the composition of planetary building blocks}",
      journal = {\aap},
         year = 2023,
        month = may,
       volume = {673},
          eid = {A117},
        pages = {A117},
          doi = {10.1051/0004-6361/202243882},
archivePrefix = {arXiv},
       eprint = {2301.05034},
 primaryClass = {astro-ph.EP},
       adsurl = {https://ui.adsabs.harvard.edu/abs/2023A&A...673A.117C}
}

@article{2023A&A...673A..17M,
	adsurl = "https://ui.adsabs.harvard.edu/abs/2023A\&A...673A..17M",
	archiveprefix = "arXiv",
	author = "{Mah}, Jingyi and {Bitsch}, Bertram",
	doi = "10.1051/0004-6361/202346021",
	eid = "A17",
	eprint = "2303.11948",
	journal = "\aap",
	month = may,
	pages = "A17",
	primaryclass = "astro-ph.EP",
	title = "{Forming super-Mercuries: Role of stellar abundances}",
	volume = "673",
	year = 2023
}

@article{2022ApJ...939L..19I,
	adsurl = "https://ui.adsabs.harvard.edu/abs/2022ApJ...939L..19I",
	archiveprefix = "arXiv",
	author = "{Izidoro}, Andr{\'e} and {Schlichting}, Hilke E. and {Isella}, Andrea and {Dasgupta}, Rajdeep and {Zimmermann}, Christian and {Bitsch}, Bertram",
	doi = "10.3847/2041-8213/ac990d",
	eid = "L19",
	eprint = "2210.05595",
	journal = "\apjl",
	month = nov,
	number = "2",
	pages = "L19",
	primaryclass = "astro-ph.EP",
	title = "{The Exoplanet Radius Valley from Gas-driven Planet Migration and Breaking of Resonant Chains}",
	volume = "939",
	year = 2022
}

@article{2023A&A...674A.178B,
	adsurl = "https://ui.adsabs.harvard.edu/abs/2023A\&A...674A.178B",
	archiveprefix = "arXiv",
	author = "{Bitsch}, Bertram and {Izidoro}, Andre",
	doi = "10.1051/0004-6361/202245040",
	eid = "A178",
	eprint = "2304.12758",
	journal = "\aap",
	month = jun,
	pages = "A178",
	primaryclass = "astro-ph.EP",
	title = "{Giants are bullies: How their growth influences systems of inner sub-Neptunes and super-Earths}",
	volume = "674",
	year = 2023
}

@article{2018MNRAS.479.4786V,
	adsurl = "https://ui.adsabs.harvard.edu/abs/2018MNRAS.479.4786V",
	archiveprefix = "arXiv",
	author = "{Van Eylen}, V. and {Agentoft}, Camilla and {Lundkvist}, M.~S. and {Kjeldsen}, H. and {Owen}, J.~E. and {Fulton}, B.~J. and {Petigura}, E. and {Snellen}, I.",
	doi = "10.1093/mnras/sty1783",
	eprint = "1710.05398",
	journal = "\mnras",
	month = oct,
	number = "4",
	pages = "4786--4795",
	primaryclass = "astro-ph.EP",
	title = "{An asteroseismic view of the radius valley: stripped cores, not born rocky}",
	volume = "479",
	year = 2018
}

@ARTICLE{2024Natur.627..501L,
       author = {{Liu}, Fan and {Ting}, Yuan-Sen and {Yong}, David and {Bitsch}, Bertram and {Karakas}, Amanda and {Murphy}, Michael T. and {Joyce}, Meridith and {Dotter}, Aaron and {Dai}, Fei},
        title = "{At least one in a dozen stars shows evidence of planetary ingestion}",
      journal = {\nat},
         year = 2024,
        month = mar,
       volume = {627},
       number = {8004},
        pages = {501-504},
          doi = {10.1038/s41586-024-07091-y},
archivePrefix = {arXiv},
       eprint = {2403.13209},
 primaryClass = {astro-ph.SR},
       adsurl = {https://ui.adsabs.harvard.edu/abs/2024Natur.627..501L}
}

@ARTICLE{2023A&A...676A..52T,
       author = {{Timmermann}, Anina and {Shan}, Yutong and {Reiners}, Ansgar and {Pack}, Andreas},
        title = "{Revisiting equilibrium condensation and rocky planet compositions. Introducing the ECCOPLANETS code}",
      journal = {\aap},
         year = 2023,
        month = aug,
       volume = {676},
          eid = {A52},
        pages = {A52},
          doi = {10.1051/0004-6361/202244850},
archivePrefix = {arXiv},
       eprint = {2307.00914},
 primaryClass = {astro-ph.EP},
       adsurl = {https://ui.adsabs.harvard.edu/abs/2023A&A...676A..52T}
}

@Article{McDonough1995,
  author  = {McDonough, William and Sun, S.S.},
  journal = {Chemical Geology},
  title   = {{The composition of the Earth}},
  year    = {1995},
  month   = {03},
  pages   = {223--253},
  doi     = {10.1016/0009-2541(94)00140-4},
}

@ARTICLE{2015A&A...577A..83D,
       author = {{Dorn}, Caroline and {Khan}, Amir and {Heng}, Kevin and {Connolly}, James A.~D. and {Alibert}, Yann and {Benz}, Willy and {Tackley}, Paul},
        title = "{Can we constrain the interior structure of rocky exoplanets from mass and radius measurements?}",
      journal = {\aap},
         year = 2015,
        month = may,
       volume = {577},
          eid = {A83},
        pages = {A83},
          doi = {10.1051/0004-6361/201424915},
archivePrefix = {arXiv},
       eprint = {1502.03605},
 primaryClass = {astro-ph.EP},
       adsurl = {https://ui.adsabs.harvard.edu/abs/2015A&A...577A..83D}
}

@ARTICLE{2022MNRAS.513.5829W,
       author = {{Wang}, H.~S. and {Quanz}, S.~P. and {Yong}, D. and {Liu}, F. and {Seidler}, F. and {Acu{\~n}a}, L. and {Mojzsis}, S.~J.},
        title = "{Detailed chemical compositions of planet-hosting stars - II. Exploration of the interiors of terrestrial-type exoplanets}",
      journal = {\mnras},
         year = 2022,
        month = jul,
       volume = {513},
       number = {4},
        pages = {5829-5846},
          doi = {10.1093/mnras/stac1119},
archivePrefix = {arXiv},
       eprint = {2204.09558},
 primaryClass = {astro-ph.EP},
       adsurl = {https://ui.adsabs.harvard.edu/abs/2022MNRAS.513.5829W}
}

@ARTICLE{2026PSJ.....7...30S,
       author = {{Schulze}, Joseph G. and {Hinkel}, Natalie R. and {Panero}, Wendy R. and {Unterborn}, Cayman T.},
        title = "{Toward Reliable Interpretations of Small-exoplanet Compositions: Comparisons and Considerations of Equations of State and Materials Used in Common Rocky Planet Models}",
      journal = {PSJ},
         year = 2026,
        month = feb,
       volume = {7},
       number = {2},
          eid = {30},
        pages = {30},
          doi = {10.3847/PSJ/ae2ea2},
archivePrefix = {arXiv},
       eprint = {2601.13176},
 primaryClass = {astro-ph.EP},
       adsurl = {https://ui.adsabs.harvard.edu/abs/2026PSJ.....7...30S}
}

@ARTICLE{2025MNRAS.542..293R,
       author = {{Rogers}, Laura K. and {Bonsor}, Amy and {Le Bourdais}, {\'E}rika and {Xu}, Siyi and {Su}, Kate Y.~L. and {Richards}, Benjamin and {Buchan}, Andrew and {Ballering}, Nicholas P. and {Brouwers}, Marc and {Dufour}, Patrick and {Kissler-Patig}, Markus and {Melis}, Carl and {Zuckerman}, Ben},
        title = "{Silicate mineralogy and bulk composition of exoplanetary material in polluted white dwarfs}",
      journal = {\mnras},
         year = 2025,
        month = sep,
       volume = {542},
       number = {1},
        pages = {293-309},
          doi = {10.1093/mnras/staf1221},
archivePrefix = {arXiv},
       eprint = {2507.16777},
 primaryClass = {astro-ph.EP},
       adsurl = {https://ui.adsabs.harvard.edu/abs/2025MNRAS.542..293R}
}

@ARTICLE{2024A&A...691A.352W,
       author = {{Williams}, J.~T. and {G{\"a}nsicke}, B.~T. and {Swan}, A. and {O'Brien}, M.~W. and {Izquierdo}, P. and {Cutolo}, A.-M. and {Cunningham}, T.},
        title = "{PEWDD: A database of white dwarfs enriched by exo-planetary material}",
      journal = {\aap},
         year = 2024,
        month = nov,
       volume = {691},
          eid = {A352},
        pages = {A352},
          doi = {10.1051/0004-6361/202450509},
archivePrefix = {arXiv},
       eprint = {2409.16046},
 primaryClass = {astro-ph.EP},
       adsurl = {https://ui.adsabs.harvard.edu/abs/2024A&A...691A.352W}
}

@ARTICLE{2019Sci...366..356D,
       author = {{Doyle}, Alexandra E. and {Young}, Edward D. and {Klein}, Beth and {Zuckerman}, Ben and {Schlichting}, Hilke E.},
        title = "{Oxygen fugacities of extrasolar rocks: Evidence for an Earth-like geochemistry of exoplanets}",
      journal = {Science},
         year = 2019,
        month = oct,
       volume = {366},
       number = {6463},
        pages = {356-359},
          doi = {10.1126/science.aax3901},
archivePrefix = {arXiv},
       eprint = {1910.12989},
 primaryClass = {astro-ph.EP},
       adsurl = {https://ui.adsabs.harvard.edu/abs/2019Sci...366..356D}
}

@ARTICLE{2001E&PSL.185...49A,
       author = {{All{\`e}gre}, Claude and {Manh{\`e}s}, G{\'e}rard and {Lewin}, {\'E}ric},
        title = "{Chemical composition of the Earth and the volatility control on planetary genetics}",
      journal = {Earth and Planetary Science Letters},
         year = 2001,
        month = feb,
       volume = {185},
       number = {1-2},
        pages = {49-69},
          doi = {10.1016/S0012-821X(00)00359-9},
       adsurl = {https://ui.adsabs.harvard.edu/abs/2001E&PSL.185...49A}
}

@ARTICLE{2021Eleme..17..235P,
       author = {{Putirka}, Keith D. and {Dorn}, Caroline and {Hinkel}, Natalie R. and {Unterborn}, Cayman T.},
        title = "{Compositional Diversity of Rocky Exoplanets}",
      journal = {Elements},
         year = 2021,
        month = aug,
       volume = {17},
       number = {4},
        pages = {235},
          doi = {10.2138/gselements.17.4.235},
archivePrefix = {arXiv},
       eprint = {2108.08383},
 primaryClass = {astro-ph.EP},
       adsurl = {https://ui.adsabs.harvard.edu/abs/2021Eleme..17..235P}
}

@ARTICLE{2024RvMG...90..259G,
       author = {{Guimond}, Claire Marie and {Wang}, Haiyang and {Seidler}, Fabian and {Sossi}, Paolo and {Mahajan}, Aprajit and {Shorttle}, Oliver},
        title = "{From Stars to Diverse Mantles, Melts, Crusts, and Atmospheres of Rocky Exoplanets}",
      journal = {Reviews in Mineralogy and Geochemistry},
         year = 2024,
        month = jul,
       volume = {90},
       number = {1},
        pages = {259-300},
          doi = {10.2138/rmg.2024.90.08},
archivePrefix = {arXiv},
       eprint = {2404.15427},
 primaryClass = {astro-ph.EP},
       adsurl = {https://ui.adsabs.harvard.edu/abs/2024RvMG...90..259G}
}

@ARTICLE{2017ApJ...840...99D,
       author = {{Dotter}, Aaron and {Conroy}, Charlie and {Cargile}, Phillip and {Asplund}, Martin},
        title = "{The Influence of Atomic Diffusion on Stellar Ages and Chemical Tagging}",
      journal = {\apj},
         year = 2017,
        month = may,
       volume = {840},
       number = {2},
          eid = {99},
        pages = {99},
          doi = {10.3847/1538-4357/aa6d10},
archivePrefix = {arXiv},
       eprint = {1704.03465},
 primaryClass = {astro-ph.SR},
       adsurl = {https://ui.adsabs.harvard.edu/abs/2017ApJ...840...99D}
}

@ARTICLE{2018MNRAS.479.3690B,
       author = {{Bitsch}, B. and {Forsberg}, R. and {Liu}, F. and {Johansen}, A.},
        title = "{Stellar abundance of binary stars: their role in determining the formation location of super-Earths and ice giants}",
      journal = {\mnras},
         year = 2018,
        month = sep,
       volume = {479},
       number = {3},
        pages = {3690-3707},
          doi = {10.1093/mnras/sty1710},
archivePrefix = {arXiv},
       eprint = {1806.10683},
 primaryClass = {astro-ph.EP},
       adsurl = {https://ui.adsabs.harvard.edu/abs/2018MNRAS.479.3690B}
}

@ARTICLE{2018A&A...618A.132K,
       author = {{Kunitomo}, Masanobu and {Guillot}, Tristan and {Ida}, Shigeru and {Takeuchi}, Taku},
        title = "{Revisiting the pre-main-sequence evolution of stars. II. Consequences of planet formation on stellar surface composition}",
      journal = {\aap},
         year = 2018,
        month = oct,
       volume = {618},
          eid = {A132},
        pages = {A132},
          doi = {10.1051/0004-6361/201833127},
archivePrefix = {arXiv},
       eprint = {1808.07396},
 primaryClass = {astro-ph.SR},
       adsurl = {https://ui.adsabs.harvard.edu/abs/2018A&A...618A.132K}
}

@ARTICLE{2023A&A...676A..87H,
       author = {{H{\"u}hn}, L.-A. and {Bitsch}, B.},
        title = "{How accretion of planet-forming disks influences stellar abundances}",
      journal = {\aap},
         year = 2023,
        month = aug,
       volume = {676},
          eid = {A87},
        pages = {A87},
          doi = {10.1051/0004-6361/202346604},
archivePrefix = {arXiv},
       eprint = {2306.16461},
 primaryClass = {astro-ph.EP},
       adsurl = {https://ui.adsabs.harvard.edu/abs/2023A&A...676A..87H}
}

@ARTICLE{2020MNRAS.493.5079B,
       author = {{Booth}, Richard A. and {Owen}, James E.},
        title = "{Fingerprints of giant planets in the composition of solar twins}",
      journal = {\mnras},
         year = 2020,
        month = apr,
       volume = {493},
       number = {4},
        pages = {5079-5088},
          doi = {10.1093/mnras/staa578},
archivePrefix = {arXiv},
       eprint = {2002.11135},
 primaryClass = {astro-ph.EP},
       adsurl = {https://ui.adsabs.harvard.edu/abs/2020MNRAS.493.5079B}
}

@ARTICLE{2024AJ....168..239D,
       author = {{Dai}, Fei and {Goldberg}, Max and {Batygin}, Konstantin and {van Saders}, Jennifer and {Chiang}, Eugene and {Choksi}, Nick and {Li}, Rixin and {Petigura}, Erik A. and {Gilbert}, Gregory J. and {Millholland}, Sarah C. and {Dai}, Yuan-Zhe and {Bouma}, Luke and {Weiss}, Lauren M. and {Winn}, Joshua N.},
        title = "{The Prevalence of Resonance Among Young, Close-in Planets}",
      journal = {\aj},
         year = 2024,
        month = dec,
       volume = {168},
       number = {6},
          eid = {239},
        pages = {239},
          doi = {10.3847/1538-3881/ad83a6},
archivePrefix = {arXiv},
       eprint = {2406.06885},
 primaryClass = {astro-ph.EP},
       adsurl = {https://ui.adsabs.harvard.edu/abs/2024AJ....168..239D}
}

@ARTICLE{2011ApJS..192....3P,
       author = {{Paxton}, Bill and {Bildsten}, Lars and {Dotter}, Aaron and {Herwig}, Falk and {Lesaffre}, Pierre and {Timmes}, Frank},
        title = "{Modules for Experiments in Stellar Astrophysics (MESA)}",
      journal = {\apjs},
         year = 2011,
        month = jan,
       volume = {192},
       number = {1},
          eid = {3},
        pages = {3},
          doi = {10.1088/0067-0049/192/1/3},
archivePrefix = {arXiv},
       eprint = {1009.1622},
 primaryClass = {astro-ph.SR},
       adsurl = {https://ui.adsabs.harvard.edu/abs/2011ApJS..192....3P}
}

@ARTICLE{2023ApJS..265...15J,
       author = {{Jermyn}, Adam S. and {Bauer}, Evan B. and {Schwab}, Josiah and {Farmer}, R. and {Ball}, Warrick H. and {Bellinger}, Earl P. and {Dotter}, Aaron and {Joyce}, Meridith and {Marchant}, Pablo and {Mombarg}, Joey S.~G. and {Wolf}, William M. and {Sunny Wong}, Tin Long and {Cinquegrana}, Giulia C. and {Farrell}, Eoin and {Smolec}, R. and {Thoul}, Anne and {Cantiello}, Matteo and {Herwig}, Falk and {Toloza}, Odette and {Bildsten}, Lars and {Townsend}, Richard H.~D. and {Timmes}, F.~X.},
        title = "{Modules for Experiments in Stellar Astrophysics (MESA): Time-dependent Convection, Energy Conservation, Automatic Differentiation, and Infrastructure}",
      journal = {\apjs},
         year = 2023,
        month = mar,
       volume = {265},
       number = {1},
          eid = {15},
        pages = {15},
          doi = {10.3847/1538-4365/acae8d},
archivePrefix = {arXiv},
       eprint = {2208.03651},
 primaryClass = {astro-ph.SR},
       adsurl = {https://ui.adsabs.harvard.edu/abs/2023ApJS..265...15J}
}

@ARTICLE{2025A&A...701A.184K,
       author = {{Kahle}, K. Angelique and {Blecic}, Jasmina and {Ashtari}, Reza and {Kreidberg}, Laura and {Kawashima}, Yui and {Cubillos}, Patricio E. and {Deming}, Drake and {Jenkins}, James S. and {Molli{\`e}re}, Paul and {Redfield}, Seth and {Tian}, Qiushi Chris and {Vines}, Jose I. and {Wilson}, David J. and {Acu{\~n}a}, Lorena and {Bitsch}, Bertram and {Brande}, Jonathan and {France}, Kevin and {Stevenson}, Kevin B. and {Crossfield}, Ian J.~M. and {Daylan}, Tansu and {Dobbs-Dixon}, Ian and {Evans-Soma}, Thomas M. and {Gapp}, Cyril and {Garc{\'\i}a Mu{\~n}oz}, Antonio and {Heng}, Kevin and {Hu}, Renyu and {Shkolnik}, Evgenya L. and {Stassun}, Keivan G. and {Teske}, Johanna},
        title = "{The SPACE Program: I. The featureless spectrum of HD 86226 c challenges sub-Neptune atmosphere trends}",
      journal = {\aap},
         year = 2025,
        month = sep,
       volume = {701},
          eid = {A184},
        pages = {A184},
          doi = {10.1051/0004-6361/202554916},
archivePrefix = {arXiv},
       eprint = {2507.13439},
 primaryClass = {astro-ph.EP},
       adsurl = {https://ui.adsabs.harvard.edu/abs/2025A&A...701A.184K}
}

@ARTICLE{2024arXiv240303325B,
       author = {{Benneke}, Bj{\"o}rn and {Roy}, Pierre-Alexis and {Coulombe}, Louis-Philippe and {Radica}, Michael and {Piaulet}, Caroline and {Ahrer}, Eva-Maria and {Pierrehumbert}, Raymond and {Krissansen-Totton}, Joshua and {Schlichting}, Hilke E. and {Hu}, Renyu and {Yang}, Jeehyun and {Christie}, Duncan and {Thorngren}, Daniel and {Young}, Edward D. and {Pelletier}, Stefan and {Knutson}, Heather A. and {Miguel}, Yamila and {Evans-Soma}, Thomas M. and {Dorn}, Caroline and {Gagnebin}, Anna and {Fortney}, Jonathan J. and {Komacek}, Thaddeus and {MacDonald}, Ryan and {Raul}, Eshan and {Cloutier}, Ryan and {Acuna}, Lorena and {Lafreni{\`e}re}, David and {Cadieux}, Charles and {Doyon}, Ren{\'e} and {Welbanks}, Luis and {Allart}, Romain},
        title = "{JWST Reveals CH$_4$, CO$_2$, and H$_2$O in a Metal-rich Miscible Atmosphere on a Two-Earth-Radius Exoplanet}",
      journal = {arXiv e-prints},
         year = 2024,
        month = mar,
          eid = {arXiv:2403.03325},
        pages = {arXiv:2403.03325},
          doi = {10.48550/arXiv.2403.03325},
archivePrefix = {arXiv},
       eprint = {2403.03325},
 primaryClass = {astro-ph.EP},
       adsurl = {https://ui.adsabs.harvard.edu/abs/2024arXiv240303325B}
}

@ARTICLE{2025arXiv250712622H,
       author = {{Hu}, Renyu and {Bello-Arufe}, Aaron and {Tokadjian}, Armen and {Yang}, Jeehyun and {Damiano}, Mario and {Roy}, Pierre-Alexis and {Coulombe}, Louis-Philippe and {Madhusudhan}, Nikku and {Constantinou}, Savvas and {Benneke}, Bj{\"o}rn},
        title = "{A water-rich interior in the temperate sub-Neptune K2-18 b revealed by JWST}",
      journal = {arXiv e-prints},
         year = 2025,
        month = jul,
          eid = {arXiv:2507.12622},
        pages = {arXiv:2507.12622},
          doi = {10.48550/arXiv.2507.12622},
archivePrefix = {arXiv},
       eprint = {2507.12622},
 primaryClass = {astro-ph.EP},
       adsurl = {https://ui.adsabs.harvard.edu/abs/2025arXiv250712622H}
}

@ARTICLE{2026arXiv260315955Z,
       author = {{Zaveri}, Urja and {Wang}, Haiyang S. and {Sossi}, Paolo A.},
        title = "{A chemical perspective on planet formation in reduced systems}",
      journal = {\aap},
         year = 2026,
        month = may,
       volume = {709},
          eid = {A223},
        pages = {A223},
          doi = {10.1051/0004-6361/202558126},
archivePrefix = {arXiv},
       eprint = {2603.15955},
 primaryClass = {astro-ph.EP},
       adsurl = {https://ui.adsabs.harvard.edu/abs/2026A&A...709A.223Z}
}

@ARTICLE{2023A&A...679L...7D,
       author = {{Danti}, C. and {Bitsch}, B. and {Mah}, J.},
        title = "{Composition of giant planets: The roles of pebbles and planetesimals}",
      journal = {\aap},
         year = 2023,
        month = nov,
       volume = {679},
          eid = {L7},
        pages = {L7},
          doi = {10.1051/0004-6361/202347501},
archivePrefix = {arXiv},
       eprint = {2310.02886},
 primaryClass = {astro-ph.EP},
       adsurl = {https://ui.adsabs.harvard.edu/abs/2023A&A...679L...7D}
}

@ARTICLE{2025NatAs...9..845E,
       author = {{Evans-Soma}, Thomas M. and {Sing}, David K. and {Barstow}, Joanna K. and {Piette}, Anjali A.~A. and {Taylor}, Jake and {Lothringer}, Joshua D. and {Reggiani}, Henrique and {Goyal}, Jayesh M. and {Ahrer}, Eva-Maria and {Mayne}, Nathan J. and {Rustamkulov}, Zafar and {Kataria}, Tiffany and {Christie}, Duncan A. and {Gapp}, Cyril and {Dong}, Jiayin and {Foreman-Mackey}, Daniel and {Hattori}, Soichiro and {Marley}, Mark S.},
        title = "{SiO and a super-stellar C/O ratio in the atmosphere of the giant exoplanet WASP-121 b}",
      journal = {Nature Astronomy},
         year = 2025,
        month = jun,
       volume = {9},
        pages = {845-861},
          doi = {10.1038/s41550-025-02513-x},
archivePrefix = {arXiv},
       eprint = {2506.01771},
 primaryClass = {astro-ph.EP},
       adsurl = {https://ui.adsabs.harvard.edu/abs/2025NatAs...9..845E}
}

@ARTICLE{2026A&A...706A..30O,
       author = {{O'Donovan}, Barry and {Bitsch}, Bertram},
        title = "{Heavy-element-enriched atmospheres and where they are born}",
      journal = {\aap},
         year = 2026,
        month = jan,
       volume = {706},
          eid = {A30},
        pages = {A30},
          doi = {10.1051/0004-6361/202556898},
archivePrefix = {arXiv},
       eprint = {2512.07944},
 primaryClass = {astro-ph.EP},
       adsurl = {https://ui.adsabs.harvard.edu/abs/2026A&A...706A..30O}
}

@ARTICLE{2024NatAs...8..463B,
       author = {{Burn}, Remo and {Mordasini}, Christoph and {Mishra}, Lokesh and {Haldemann}, Jonas and {Venturini}, Julia and {Emsenhuber}, Alexandre and {Henning}, Thomas},
        title = "{A radius valley between migrated steam worlds and evaporated rocky cores}",
      journal = {Nature Astronomy},
         year = 2024,
        month = apr,
       volume = {8},
        pages = {463-471},
          doi = {10.1038/s41550-023-02183-7},
archivePrefix = {arXiv},
       eprint = {2401.04380},
 primaryClass = {astro-ph.EP},
       adsurl = {https://ui.adsabs.harvard.edu/abs/2024NatAs...8..463B}
}

@ARTICLE{2024A&A...692A.246B,
       author = {{Bitsch}, Bertram and {Izidoro}, Andre},
        title = "{Diversity of disc viscosities can explain the period ratios of resonant and non-resonant systems of hot super-Earths and mini-Neptunes}",
      journal = {\aap},
         year = 2024,
        month = dec,
       volume = {692},
          eid = {A246},
        pages = {A246},
          doi = {10.1051/0004-6361/202451017},
archivePrefix = {arXiv},
       eprint = {2411.11452},
 primaryClass = {astro-ph.EP},
       adsurl = {https://ui.adsabs.harvard.edu/abs/2024A&A...692A.246B}
}

@ARTICLE{2021NatAs...5.1163S,
       author = {{Spina}, Lorenzo and {Sharma}, Parth and {Mel{\'e}ndez}, Jorge and {Bedell}, Megan and {Casey}, Andrew R. and {Carlos}, Mar{\'\i}lia and {Franciosini}, Elena and {Vallenari}, Antonella},
        title = "{Chemical evidence for planetary ingestion in a quarter of Sun-like stars}",
      journal = {Nature Astronomy},
         year = 2021,
        month = nov,
       volume = {5},
        pages = {1163-1169},
          doi = {10.1038/s41550-021-01451-8},
archivePrefix = {arXiv},
       eprint = {2108.12040},
 primaryClass = {astro-ph.SR},
       adsurl = {https://ui.adsabs.harvard.edu/abs/2021NatAs...5.1163S}
}

@ARTICLE{2023MNRAS.526.2181Y,
       author = {{Yong}, David and {Liu}, Fan and {Ting}, Yuan-Sen and {Joyce}, Meridith and {Bitsch}, Bertram and {Dai}, Fei and {Dotter}, Aaron and {Karakas}, Amanda I. and {Murphy}, Michael T.},
        title = "{C3PO: towards a complete census of co-moving pairs of stars - I. High precision stellar parameters for 250 stars}",
      journal = {\mnras},
         year = 2023,
        month = dec,
       volume = {526},
       number = {2},
        pages = {2181-2195},
          doi = {10.1093/mnras/stad2679},
archivePrefix = {arXiv},
       eprint = {2309.01546},
 primaryClass = {astro-ph.SR},
       adsurl = {https://ui.adsabs.harvard.edu/abs/2023MNRAS.526.2181Y}
}

@ARTICLE{2022A&A...660A..85J,
       author = {{Jorge}, D.~M. and {Kamp}, I.~E.~E. and {Waters}, L.~B.~F.~M. and {Woitke}, P. and {Spaargaren}, R.~J.},
        title = "{Forming Planets Around Stars With Non-Solar Elemental Composition}",
      journal = {\aap},
         year = 2022,
        month = apr,
       volume = {660},
          eid = {A85},
        pages = {A85},
          doi = {10.1051/0004-6361/202142738},
archivePrefix = {arXiv},
       eprint = {2202.13920},
 primaryClass = {astro-ph.EP},
       adsurl = {https://ui.adsabs.harvard.edu/abs/2022A&A...660A..85J}
}

@ARTICLE{2025A&A...703A.218S,
       author = {{Spaargaren}, Rob J. and {Herbort}, Oliver and {Wang}, Haiyang S. and {Mojzsis}, Stephen J. and {Sossi}, Paolo},
        title = "{Proto-planetary disk composition-dependent element volatility in the context of rocky planet formation}",
      journal = {\aap},
         year = 2025,
        month = nov,
       volume = {703},
          eid = {A218},
        pages = {A218},
          doi = {10.1051/0004-6361/202556011},
archivePrefix = {arXiv},
       eprint = {2509.03724},
 primaryClass = {astro-ph.EP},
       adsurl = {https://ui.adsabs.harvard.edu/abs/2025A&A...703A.218S}
}

@ARTICLE{2021JGRE..12606639B,
       author = {{Bean}, Jacob L. and {Raymond}, Sean N. and {Owen}, James E.},
        title = "{The Nature and Origins of Sub-Neptune Size Planets}",
      journal = {Journal of Geophysical Research (Planets)},
         year = 2021,
        month = jan,
       volume = {126},
       number = {1},
          eid = {e06639},
        pages = {e06639},
          doi = {10.1029/2020JE006639},
archivePrefix = {arXiv},
       eprint = {2010.11867},
 primaryClass = {astro-ph.EP},
       adsurl = {https://ui.adsabs.harvard.edu/abs/2021JGRE..12606639B}
}

@ARTICLE{2025ApJ...991L..16W,
       author = {{Werlen}, Aaron and {Dorn}, Caroline and {Burn}, Remo and {Schlichting}, Hilke E. and {Grimm}, Simon L. and {Young}, Edward D.},
        title = "{Sub-Neptunes Are Drier than They Seem: Rethinking the Origins of Water-rich Worlds}",
      journal = {\apjl},
         year = 2025,
        month = sep,
       volume = {991},
       number = {1},
          eid = {L16},
        pages = {L16},
          doi = {10.3847/2041-8213/adff73},
archivePrefix = {arXiv},
       eprint = {2507.00765},
 primaryClass = {astro-ph.EP},
       adsurl = {https://ui.adsabs.harvard.edu/abs/2025ApJ...991L..16W}
}

@ARTICLE{2026arXiv260415891N,
       author = {{Nicholls}, Harrison and {Shorttle}, Oliver and {Lichtenberg}, Tim and {Pascal}, Flavia},
        title = "{Beyond the mass-radius plane: integrated radiative-convective and interior structure simulations of the exoplanet continuum}",
      journal = {\mnras},
         year = 2026,
        month = sep,
       volume = {551},
       number = {2},
          eid = {stag1489},
        pages = {stag1489},
          doi = {10.1093/mnras/stag1489},
archivePrefix = {arXiv},
       eprint = {2604.15891},
 primaryClass = {astro-ph.EP},
       adsurl = {https://ui.adsabs.harvard.edu/abs/2026MNRAS.551g1489N}
}

@ARTICLE{2010Icar..205..321B,
       author = {{Bond}, Jade C. and {Lauretta}, Dante S. and {O'Brien}, David P.},
        title = "{Making the Earth: Combining dynamics and chemistry in the Solar System}",
      journal = {\icarus},
         year = 2010,
        month = feb,
       volume = {205},
       number = {2},
        pages = {321-337},
          doi = {10.1016/j.icarus.2009.07.037},
archivePrefix = {arXiv},
       eprint = {0911.0426},
 primaryClass = {astro-ph.EP},
       adsurl = {https://ui.adsabs.harvard.edu/abs/2010Icar..205..321B}
}

@ARTICLE{2025A&A...694A.303H,
       author = {{Hatalova}, P. and {Brodholt}, J.~P. and {Brasser}, R. and {Shan}, Y. and {Werner}, S.~C.},
        title = "{The compositional diversity of rocky exoplanets around K-dwarf stars}",
      journal = {\aap},
         year = 2025,
        month = feb,
       volume = {694},
          eid = {A303},
        pages = {A303},
          doi = {10.1051/0004-6361/202451070},
       adsurl = {https://ui.adsabs.harvard.edu/abs/2025A&A...694A.303H}
}

@ARTICLE{2010A&A...521A..33R,
       author = {{Ram{\'\i}rez}, I. and {Asplund}, M. and {Baumann}, P. and {Mel{\'e}ndez}, J. and {Bensby}, T.},
        title = "{A possible signature of terrestrial planet formation in the chemical composition of solar analogs}",
      journal = {\aap},
         year = 2010,
        month = oct,
       volume = {521},
          eid = {A33},
        pages = {A33},
          doi = {10.1051/0004-6361/201014456},
archivePrefix = {arXiv},
       eprint = {1008.3161},
 primaryClass = {astro-ph.SR},
       adsurl = {https://ui.adsabs.harvard.edu/abs/2010A&A...521A..33R}
}

@ARTICLE{2018ApJ...865...68B,
       author = {{Bedell}, Megan and {Bean}, Jacob L. and {Mel{\'e}ndez}, Jorge and {Spina}, Lorenzo and {Ram{\'\i}rez}, Ivan and {Asplund}, Martin and {Alves-Brito}, Alan and {dos Santos}, Leonardo and {Dreizler}, Stefan and {Yong}, David and {Monroe}, TalaWanda and {Casagrande}, Luca},
        title = "{The Chemical Homogeneity of Sun-like Stars in the Solar Neighborhood}",
      journal = {\apj},
         year = 2018,
        month = sep,
       volume = {865},
       number = {1},
          eid = {68},
        pages = {68},
          doi = {10.3847/1538-4357/aad908},
archivePrefix = {arXiv},
       eprint = {1802.02576},
 primaryClass = {astro-ph.SR},
       adsurl = {https://ui.adsabs.harvard.edu/abs/2018ApJ...865...68B}
}

@ARTICLE{2025ExA....59...26R,
       author = {{Rauer}, Heike and {Aerts}, Conny and {Cabrera}, Juan and {Deleuil}, Magali and {Erikson}, Anders and {Gizon}, Laurent and {Goupil}, Mariejo and {Heras}, Ana and {Walloschek}, Thomas and {Lorenzo-Alvarez}, Jose and {Marliani}, Filippo and {Martin-Garcia}, C{\'e}sar and {Mas-Hesse}, J. Miguel and {O'Rourke}, Laurence and {Osborn}, Hugh and {Pagano}, Isabella and {Piotto}, Giampaolo and {Pollacco}, Don and {Ragazzoni}, Roberto and {Ramsay}, Gavin and {Udry}, St{\'e}phane and {Appourchaux}, Thierry and {Benz}, Willy and {Brandeker}, Alexis and {G{\"u}del}, Manuel and {Janot-Pacheco}, Eduardo and {Kabath}, Petr and {Kjeldsen}, Hans and {Min}, Michiel and {Santos}, Nuno and {Smith}, Alan and {Suarez}, Juan-Carlos and {Werner}, Stephanie C. and {Aboudan}, Alessio and {Abreu}, Manuel and {Acu{\~n}a}, Lorena and {Adams}, Moritz and {Adibekyan}, Vardan and {Affer}, Laura and {Agneray}, Fran{\c{c}}ois and {Agnor}, Craig and {Aguirre B{\o}rsen-Koch}, Victor and {Ahmed}, Saad and {Aigrain}, Suzanne and {Al-Bahlawan}, Ashraf and {Alcacera Gil}, Ma de los Angeles and {Alei}, Eleonora and {Alencar}, Silvia and {Alexander}, Richard and {Alfonso-Garz{\'o}n}, Julia and {Alibert}, Yann and {Allende Prieto}, Carlos and {Almeida}, Leonardo and {Alonso Sobrino}, Roi and {Altavilla}, Giuseppe and {Althaus}, Christian and {Alvarez Trujillo}, Luis Alonso and {Amarsi}, Anish and {Ammler-von Eiff}, Matthias and {Am{\^o}res}, Eduardo and {Andrade}, Laerte and {Antoniadis-Karnavas}, Alexandros and {Ant{\'o}nio}, Carlos and {Aparicio del Moral}, Beatriz and {Appolloni}, Matteo and {Arena}, Claudio and {Armstrong}, David and {Aroca Aliaga}, Jose and {Asplund}, Martin and {Audenaert}, Jeroen and {Auricchio}, Natalia and {Avelino}, Pedro and {Baeke}, Ann and {Bailli{\'e}}, Kevin and {Balado}, Ana and {Ballber Balaguer{\'o}}, Pau and {Balestra}, Andrea and {Ball}, Warrick and {Ballans}, Herve and {Ballot}, Jerome and {Barban}, Caroline and {Barbary}, Ga{\"e}le and {Barbieri}, Mauro and {Barcel{\'o} Forteza}, Sebasti{\`a} and {Barker}, Adrian and {Barklem}, Paul and {Barnes}, Sydney and {Barrado Navascues}, David and {Barragan}, Oscar and {Baruteau}, Cl{\'e}ment and {Basu}, Sarbani and {Baudin}, Frederic and {Baumeister}, Philipp and {Bayliss}, Daniel and {Bazot}, Michael and {Beck}, Paul G. and {Belkacem}, Kevin and {Bellinger}, Earl and {Benatti}, Serena and {Benomar}, Othman and {B{\'e}rard}, Diane and {Bergemann}, Maria and {Bergomi}, Maria and {Bernardo}, Pierre and {Biazzo}, Katia and {Bignamini}, Andrea and {Bigot}, Lionel and {Billot}, Nicolas and {Binet}, Martin and {Biondi}, David and {Biondi}, Federico and {Birch}, Aaron C. and {Bitsch}, Bertram and {Bluhm Ceballos}, Paz Victoria and {B{\'o}di}, Attila and {Bogn{\'a}r}, Zs{\'o}fia and {Boisse}, Isabelle and {Bolmont}, Emeline and {Bonanno}, Alfio and {Bonavita}, Mariangela and {Bonfanti}, Andrea and {Bonfils}, Xavier and {Bonito}, Rosaria and {Bonomo}, Aldo Stefano and {B{\"o}rner}, Anko and {Boro Saikia}, Sudeshna and {Borreguero Mart{\'\i}n}, Elisa and {Borsa}, Francesco and {Borsato}, Luca and {Bossini}, Diego and {Bouchy}, Francois and {Bou{\'e}}, Gwena{\"e}l and {Boufleur}, Rodrigo and {Boumier}, Patrick and {Bourrier}, Vincent and {Bowman}, Dominic M. and {Bozzo}, Enrico and {Bradley}, Louisa and {Bray}, John and {Bressan}, Alessandro and {Breton}, Sylvain and {Brienza}, Daniele and {Brito}, Ana and {Brogi}, Matteo and {Brown}, Beverly and {Brown}, David J.~A. and {Brun}, Allan Sacha and {Bruno}, Giovanni and {Bruns}, Michael and {Buchhave}, Lars A. and {Bugnet}, Lisa and {Buldgen}, Ga{\"e}l and {Burgess}, Patrick and {Busatta}, Andrea and {Busso}, Giorgia and {Buzasi}, Derek and {Caballero}, Jos{\'e} A. and {Cabral}, Alexandre and {Cabrero Gomez}, Juan-Francisco and {Calderone}, Flavia and {Cameron}, Robert and {Cameron}, Andrew and {Campante}, Tiago and {Campos Gestal}, N{\'e}stor and {Canto Martins}, Bruno Leonardo and {Cara}, Christophe and {Carone}, Ludmila and {Carrasco}, Josep Manel and {Casagrande}, Luca and {Casewell}, Sarah L. and {Cassisi}, Santi and {Castellani}, Marco and {Castro}, Matthieu and {Catala}, Claude and {Catal{\'a}n Fern{\'a}ndez}, Irene and {Catelan}, M{\'a}rcio and {Cegla}, Heather and {Cerruti}, Chiara and {Cessa}, Virginie and {Chadid}, Merieme and {Chaplin}, William and {Charpinet}, Stephane and {Chiappini}, Cristina and {Chiarucci}, Simone and {Chiavassa}, Andrea and {Chinellato}, Simonetta and {Chirulli}, Giovanni and {Christensen-Dalsgaard}, J{\o}rgen and {Church}, Ross and {Claret}, Antonio and {Clarke}, Cathie and {Claudi}, Riccardo and {Clermont}, Lionel and {Coelho}, Hugo and {Coelho}, Joao and {Cogato}, Fabrizio and {Colom{\'e}}, Josep and {Condamin}, Mathieu and {Conde Garc{\'\i}a}, Fernando and {Conseil}, Simon},
        title = "{The PLATO mission}",
      journal = {Experimental Astronomy},
         year = 2025,
        month = jun,
       volume = {59},
       number = {3},
          eid = {26},
        pages = {26},
          doi = {10.1007/s10686-025-09985-9},
archivePrefix = {arXiv},
       eprint = {2406.05447},
 primaryClass = {astro-ph.IM},
       adsurl = {https://ui.adsabs.harvard.edu/abs/2025ExA....59...26R}
}

@ARTICLE{2026arXiv260701699B,
       author = {{Babatsikos}, Mia and {Liu}, Fan and {Karakas}, Amanda and {Mandel}, Ilya and {Passenger}, Lachlan and {Bedell}, Megan and {Wang}, Sharon X. and {Cheng}, Zimo},
        title = "{The Sun's Chemical Peculiarity: Disentangling Galactic Chemical Evolution and Planetary Engulfment in Solar Twins}",
      journal = {\apj},
         year = 2026,
        month = oct,
       volume = {1010},
       number = {1},
          eid = {96},
        pages = {96},
          doi = {10.3847/1538-4357/aea561},
       adsurl = {https://ui.adsabs.harvard.edu/abs/2026ApJ..1010...96B}
}

@ARTICLE{2024A&A...682L..23S,
       author = {{Saffe}, C. and {Miquelarena}, P. and {Alacoria}, J. and {Martioli}, E. and {Flores}, M. and {Jaque Arancibia}, M. and {Angeloni}, R. and {Jofr{\'e}}, E. and {Yana Galarza}, J. and {Gonz{\'a}lez}, E. and {Collado}, A.},
        title = "{Disentangling the origin of chemical differences using GHOST}",
      journal = {\aap},
         year = 2024,
        month = feb,
       volume = {682},
          eid = {L23},
        pages = {L23},
          doi = {10.1051/0004-6361/202449263},
       adsurl = {https://ui.adsabs.harvard.edu/abs/2024A&A...682L..23S}
}

@ARTICLE{2025A&A...693A..64A,
       author = {{Aguilera-G{\'o}mez}, Claudia and {Rogers}, Laura K. and {Bonsor}, Amy and {Jofr{\'e}}, Paula and {Blouin}, Simon and {Shorttle}, Oliver and {Buchan}, Andrew M. and {Li}, Yuqi and {Xu}, Siyi},
        title = "{Host star and exoplanet composition: Polluted white dwarf reveals depletion of moderately refractory elements in planetary material}",
      journal = {\aap},
         year = 2025,
        month = jan,
       volume = {693},
          eid = {A64},
        pages = {A64},
          doi = {10.1051/0004-6361/202451621},
archivePrefix = {arXiv},
       eprint = {2411.03454},
 primaryClass = {astro-ph.SR},
       adsurl = {https://ui.adsabs.harvard.edu/abs/2025A&A...693A..64A}
}

@ARTICLE{2000ApJ...529L..45C,
       author = {{Charbonneau}, David and {Brown}, Timothy M. and {Latham}, David W. and {Mayor}, Michel},
        title = "{Detection of Planetary Transits Across a Sun-like Star}",
      journal = {\apjl},
         year = 2000,
        month = jan,
       volume = {529},
       number = {1},
        pages = {L45-L48},
          doi = {10.1086/312457},
archivePrefix = {arXiv},
       eprint = {astro-ph/9911436},
 primaryClass = {astro-ph},
       adsurl = {https://ui.adsabs.harvard.edu/abs/2000ApJ...529L..45C}
}

@ARTICLE{2026arXiv260824407W,
       author = {{Wang}, Haiyang S. and {Johansen}, Anders and {Xu}, Ziyan and {Steinmeyer}, Marie-Luise and {Lambrechts}, Michiel and {van Kooten}, Elishevah and {Yang}, Chao-Chin and {Zhu}, Zhaohuan and {Lauretta}, Dante S. and {Bizzarro}, Martin},
        title = "{Volatile depletion in rocky planets as a chemical fingerprint of hybrid accretion}",
      journal = {Nature Astronomy},
         year = 2026,
        month = sep,
          doi = {10.1038/s41550-026-02984-6},
       adsurl = {https://ui.adsabs.harvard.edu/abs/2026NatAs.tmp..216W}
}

 \clearpage

\begin{appendix}

\section{Solar abundances}
\label{ap:solar}

We list in table~\ref{tab:solar_values} the solar abundances used as a reference point in this work \citep{2009ARA&A..47..481A}.

\begin{table}
 \centering
  \caption{Elemental number ratios used in our model, corresponding to the abundance of element X compared to hydrogen in the solar photosphere \citep{2009ARA&A..47..481A}.}
 \label{tab:solar_values}
 \begin{tabular}{c c}
 \hline
 \hline
 Species (X) & Abundance \\ \hline
 He/H & 0.085\\
 O/H & $4.90\times 10^{-4}$ \\
 C/H & $2.69\times 10^{-4}$ \\
 Mg/H & $3.98\times 10^{-5}$ \\
 Si/H & $3.24\times 10^{-5}$ \\
 Fe/H & $3.16\times 10^{-5}$ \\
 S/H & $1.32\times 10^{-5}$ \\
 \hline
 \end{tabular}

\end{table}

\section{Stellar abundance measurements}
\label{ap:abundance}

\begin{table*}
\centering
\caption{Stellar parameters of the different stars for our 7 co-natal pairs. }
\begin{tabular}{lllllll}
Pair & Stellar name & Gaia DR3 ID & Mass & T$_{\rm eff}$ & logg & Magnitude \\ \hline
69 Star A & HD 24062 & 4666907551119833984 & 1.22 & 6000 & 4.25 & 7.28 \\
69 Star B & HD24085 & 4666907516760096512 & 1.18 & 6056 & 4.32 & 7.43 \\ \hline

74 Star A & HD 25927B & 4857866023347844608 & 1.07 & 6036 & 4.49 & 9.17 \\
74 Star B & HD 25927A & 4857866019051550976 & 1.20 & 6318 & 4.42 & 8.67 \\ \hline

77 Star A & CD-80 348 & 5195685713943687168 & 1.11 & 6124 & 4.45 & 9.72  \\
77 Star B & CD-80 346 & 5195685713943683072 & 1.11 & 6169 & 4.39 & 9.61 \\ \hline

79 Star A & HD 78964 & 5296896240321814144 & 0.83 & 5138 & 4.39 & 9.63  \\
79 Star B & HD 78977 & 5296896205962076416 & 0.92 & 5355 & 4.47 & 9.29 \\ \hline

112 Star A & TYC 8832-243-1 & 6497943781186827264 & 1.09 & 5800 & 4.52 & 10.07 \\
112 Star B & TYC 8832-1422-1 & 6497943781187251072 & 0.99 & 5571 & 4.53 & 10.48 \\ \hline

116 Star A & CD-52 9070 & 6644929416166727680 & 1.13 & 5892 & 4.30 & 9.36  \\
116 Star B & CD-52 9069 & 6644929656684912512 & 1.10 & 5875 & 4.34 & 9.50 \\ \hline

124 Star A & HD 185726 & 6868438517820721152 & 1.13 & 6271 & 4.20 & 8.84  \\
124 Star B & HD 185689 & 6868432504866186368 & 1.09 & 6132 & 4.36 & 9.42 \\ \hline
  \hline
\end{tabular}
\tablefoot{The corresponding stellar abundances can be found in table~\ref{tab:stellarabundances}.}
\label{tab:stellarparameters}
\end{table*}

\begin{figure*}
 \centering
 \includegraphics[scale=0.6]{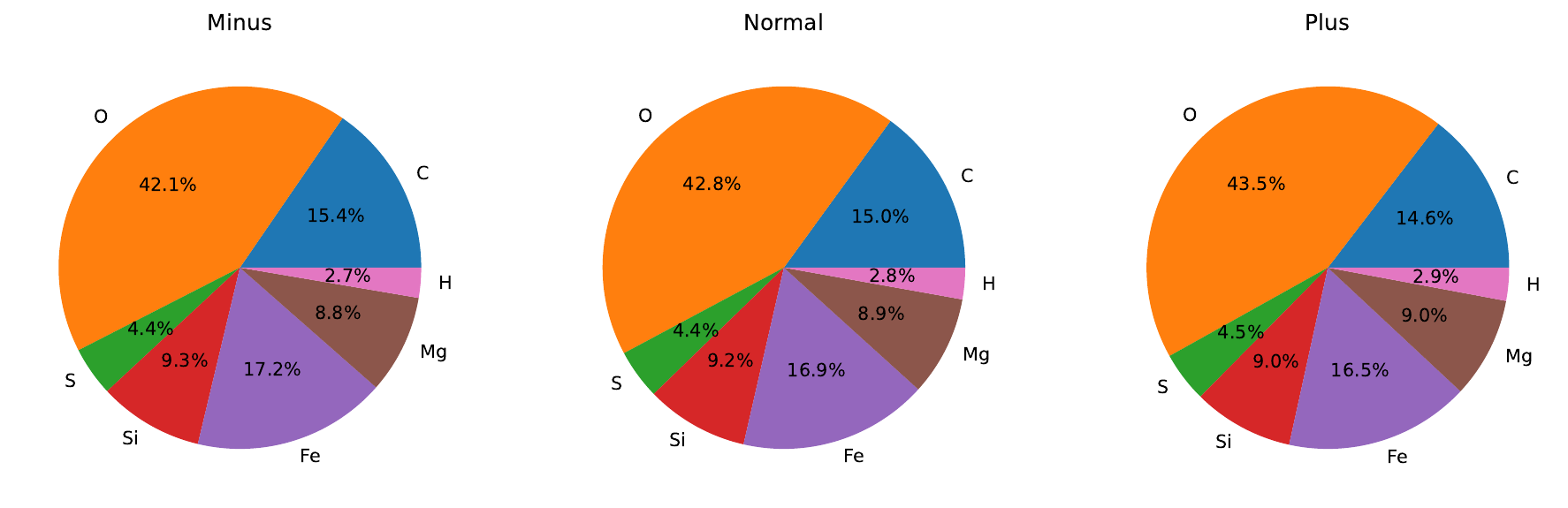} \quad
 \includegraphics[scale=0.6]{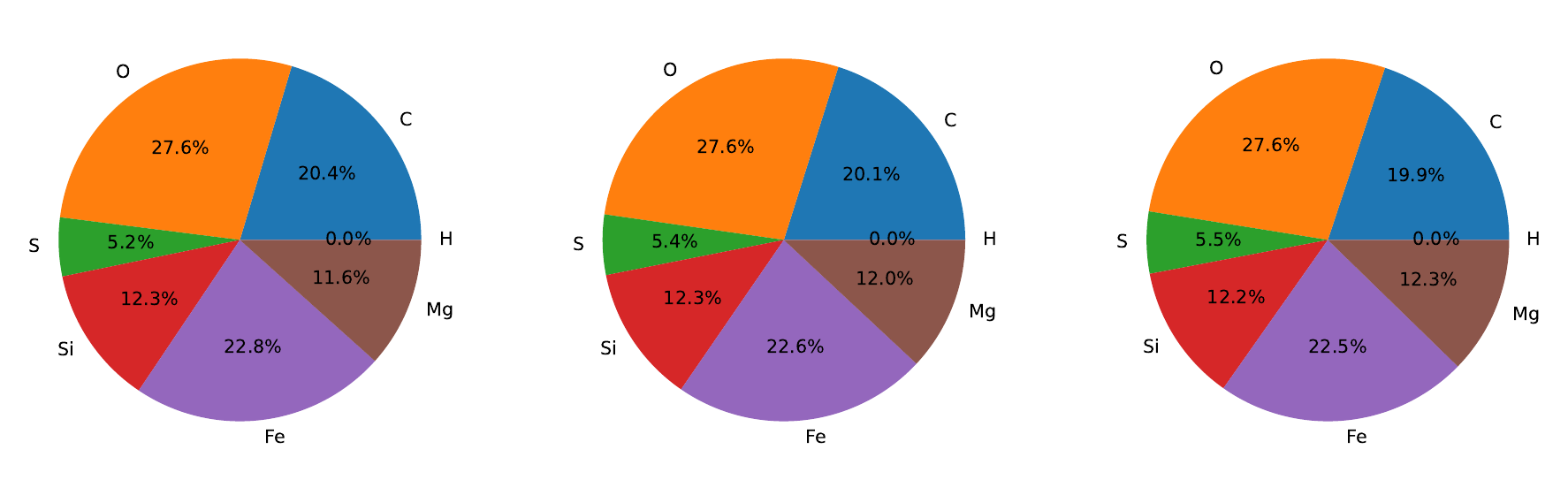}
 \caption{Mass fraction of planets formed at 140K (top) and 300K (bottom) around Star B in pair 69. The composition of the planet is determined by the relative abundances of the elements the planet accreted. From left to right we change the stellar abundances according to their error bars in table~\ref{tab:stellarabundances}, where ``Minus'' corresponds to [X/H]$-\sigma$, ``Normal'' to [X/H], and ``Plus'' to [X/H]$+\sigma$.
   \label{fig:Massfraction}
   }
\end{figure*}

\begin{table*}
\caption{\normalsize Stellar abundances of the stars in each stellar pairs as well as their abundance differences determined by Star A minus Star B.}
\label{tab:stellarabundances}
\centering {\small
\begin{tabular}{llllllll}
Pair & Object & [Fe/H] & [Mg/H] & [Si/H] & [S/H] & [O/H] & [C/H]\\ \hline
69 & $\Delta$ [X/H] & $-0.061 \pm 0.0058$ & $-0.064 \pm 0.0167$ & $-0.052 \pm 0.0027$ & $-0.007 \pm 0.0251$ & $-0.001\pm 0.0063$ & $0.005 \pm 0.0025$ \\
 & Star A & $0.224 \pm 0.014$ & $0.197 \pm 0.0265$ & $0.241 \pm 0.0133$ & $0.214 \pm 0.026$ & $0.044 \pm 0.0272$ & $0.069 \pm 0.0108$\\
 & Star B & $0.164 \pm 0.0111$ & $0.149 \pm 0.0259$ & $0.189 \pm 0.0121$ & $0.207 \pm 0.0256$ & $0.04 \pm 0.0235$ & $0.074 \pm 0.0087$ \\ \hline

 74 & $\Delta$ [X/H] & $0.075\pm 0.0123$ & $0.072\pm 0.0261$ & $0.078\pm 0.0123$ & $0.032\pm 0.0259$ & - & $0.005 \pm 0.0268$ \\
  & Star A & -0.156 $\pm$ 0.0155 & -0.181 $\pm$ 0.0113 & -0.173 $\pm$ 0.0113 & -0.129 $\pm$ 0.026 & - & -0.226 $\pm$ 0.0263\\
  & Star B & -0.082 $\pm$ 0.0216 & -0.109 $\pm$ 0.028 & -0.095 $\pm$ 0.0168 & -0.097 $\pm$ 0.0266 & - & -0.222 $\pm$ 0.0286 \\
 \hline

 77 & $\Delta$ [X/H] & $-0.0459 \pm 0.0141$ & $-0.051\pm0.0263$ & $-0.0655\pm 0.0139$ & $-0.037\pm 0.0258$ & - & $0.0245\pm 0.0231$ \\
  & Star A & -0.1225 $\pm$ 0.017 & -0.1185 $\pm$ 0.0278 & -0.089 $\pm$ 0.0163 & -0.1395 $\pm$ 0.0271 & - & -0.2385 $\pm$ 0.0276 \\
  & Star B & -0.171 $\pm$ 0.0191 & -0.1695 $\pm$ 0.0275 & -0.1555 $\pm$ 0.0147 & -0.1765 $\pm$ 0.0269 & - & -0.22 $\pm$ 0.028 \\
 \hline

 79 & $\Delta$ [X/H] & 0.036 $\pm$ 0.059 & 0.044 $\pm$ 0.0257 & 0.014 $\pm$ 0.0094 & - & - & -0.124 $\pm$ 0.0272 \\
  & Star A & 0.1 $\pm$ 0.0169 & 0.04 $\pm$ 0.0294 & 0.132 $\pm$ 0.0187 & - & - & 0.062 $\pm$ 0.00359 \\
  & Star B & 0.136 $\pm$ 0.0123 & 0.084 $\pm$ 0.0263 & 0.144 $\pm$ 0.0129 & - & - & -0.062 $\pm$ 0.0272 \\
 \hline

 112 & $\Delta$ [X/H] & $-0.087 \pm 0.0076$ & $-0.053 \pm 0.0257$ & $-0.079 \pm 0.0087$ & $0.173 \pm 0.0268$ & $0.031 \pm 0.0305$ & $-0.02 \pm 0.0269$ \\
  & Star A & $0.173 \pm 0.0088$ & $0.123 \pm 0.0262$ & $0.134 \pm 0.006$ & $0.128 \pm 0.0273$ & $0.03 \pm 0.017$ & $-0.09 \pm 0.0278$ \\
  & Star B & $0.088 \pm 0.0107$ & $0.07 \pm 0.0264$ & $0.055 \pm 0.0078$ & $0.301 \pm 0.0283$ & $0.089 \pm 0.0197$ & $-0.11 \pm 0.0286$ \\ \hline

  116 & $\Delta$ [X/H] & -0.044 $\pm$ 0.0088 & -0.015 $\pm$ 0.0257 & -0.049 $\pm$ 0.0061 & -0.051 $\pm$ 0.0256 & 0.011 $\pm$ 0.0177 & 0.108 $\pm$ 0.0109\\
   & Star A & 0.168 $\pm$ 0.0102 & 0.177 $\pm$ 0.026 & 0.186 $\pm$ 0.0096 & 0.194 $\pm$ 0.0265 & 0.066 $\pm$ 0.0163 & 0.047 $\pm$ 0.0756 \\
   & Star B & 0.126 $\pm$ 0.0088 & 0.162 $\pm$ 0.0259 & 0.137 $\pm$ 0.0077 & 0.143 $\pm$ 0.026 & 0.089 $\pm$ 0.0132 & 0.155 $\pm$ 0.0665 \\
  \hline

  124 & $\Delta$ [X/H] & 0.158 $\pm$ 0.0091 & 0.197 $\pm$ 0.0083 & 0.129 $\pm$ 0.0088 & 0.026 $\pm$ 0.0257 & 0.052 $\pm$ 0.0114 & 0.006 $\pm$ 0.0197 \\
   & Star A & -0.365 $\pm$ 0.0191 & -0.29 $\pm$ 0.029 & -0.274 $\pm$ 0.0131 & -0.302 $\pm$ 0.0277 & -0.249 $\pm$ 0.0333 & -0.33 $\pm$ 0.0302 \\
   & Star B & -0.208 $\pm$ 0.0117 & -0.094 $\pm$ 0.0266 & -0.145 $\pm$ 0.0071 & -0.276 $\pm$ 0.027 & -0.195 $\pm$ 0.0275 & -0.34 $\pm$ 0.0166 \\
  \hline
\end{tabular} }

\end{table*}

\begin{figure}
 \centering\includegraphics[scale=0.6]{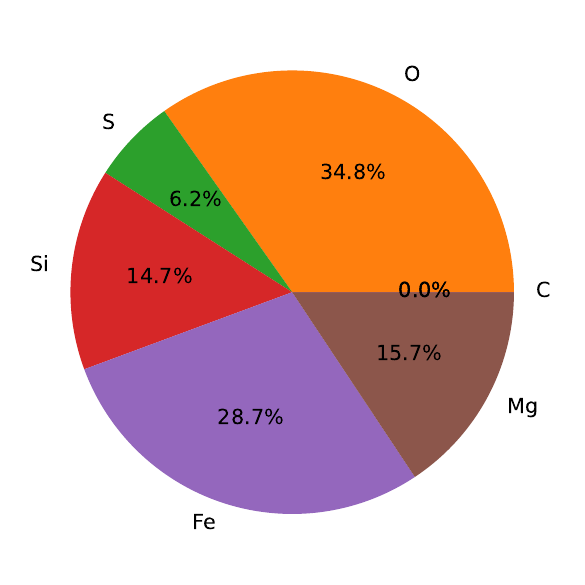}
 \caption{Mass fraction of a rocky planet formed in a disc with solar composition. The planet formed at 650K, so our model would not allow that carbon is present in the planet, similar to Earth.
   \label{fig:Massfractionsolar}
   }
\end{figure}

We show in table~\ref{tab:stellarabundances} the measured stellar abundances of the individual stars for each stellar pair as well as the differences of the stellar abundances determined by Star B minus Star A, while we show the stellar parameters in table~\ref{tab:stellarparameters}. The stellar abundances have been determined by \citet{2023MNRAS.526.2181Y, 2024Natur.627..501L}. For simplicity we use the iron abundances determined by the FeI line. The technical aspect of the measurements and the measurements themselves are discussed in \cite{2023MNRAS.526.2181Y}. Essentially, Star A's parameters and abundances were determined relative to the Sun (as absolute values). Star B's parameters and abundances were then determined relative to Star A (line-to-line). This significantly suppresses the systematic errors in the abundances measurements.

In order to understand the influence of the errors of the measured stellar abundances, we show in Fig.~\ref{fig:Massfraction} the mass fractions of planets formed at 140K (so beyond the water ice line) including the extreme ranges of the stellar abundance measurements. While the error bars indeed influence the relative abundances of all the planet forming elements, the influence of the errors in the stellar abundances is very minimal as the abundances deviate by around 1\% at maximum. In addition, these variations are small compared to the changes we expect if the planet formed at 300K (bottom in Fig.~\ref{fig:Massfraction}), so interior to the water ice line, where the planet does not accrete water ice any more. In this case, the planet's oxygen abundance decreases and the planet does not contain hydrogen any more due to fact that the planet does not accrete water. Consequently, the abundances of all other elements increases significantly.

This shows that including a detailed error evaluation in determining the masses and compositions of the planets that the stars have swallowed to explain their abundances differences will be minimal and do not influence the general conclusions in a qualitative way. Instead, the abundance differences are driven by the different formation locations in the disc.

We show in Fig.~\ref{fig:Massfractionsolar} the composition our model would predict for a rocky planet formed at 650K (so without carbon) in a disc with solar composition \citep{2009ARA&A..47..481A}. The resulting chemical composition is slightly different compared to the Earth \citep{McDonough1995}, because we only include 6 elements in our model. Nevertheless, this illustrates that the composition of (rocky) exo-planets can be different compared to the terrestrial planets in the solar system (see also \citealt{2021Sci...374..330A}).

\section{Influence of the stellar convective zone}
\label{ap:convection}

\begin{figure}
 \centering
 \includegraphics[scale=0.5]{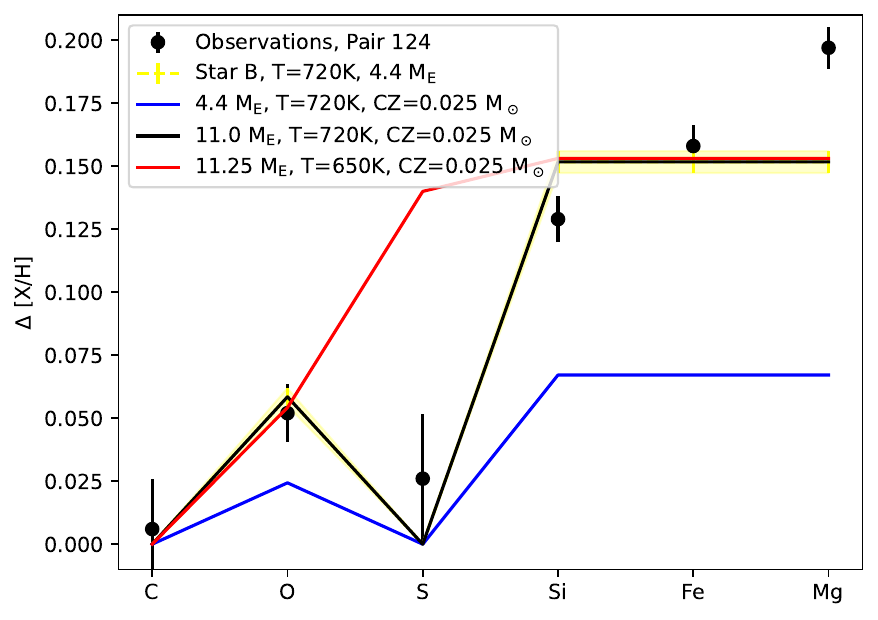}
 \caption{Abundance differences of stellar pair 124, as well as the calculated change of the stellar abundance of star B for planets of different masses and for different stellar convective zones. The yellow curve corresponds to our nominal fit with a convective zone corresponding to 0.01 solar masses (Fig.~\ref{fig:pairs}).
   \label{fig:convect}
   }
\end{figure}

In order to test if the mass of the stellar convective zone influences our results, we show in Fig.~\ref{fig:convect} the stellar abundance differences caused by ingested planets of different masses for stars with varying stellar convective zones. The planetary material always stems from close to the star, with formation temperatures of 650 and 720 K. In the nominal case (Fig.~\ref{fig:pairs}), the stellar convective zone corresponds to 0.01 solar masses. If the convective zone would be larger (e.g. 0.025 solar masses), the same 4.4 Earth mass planet would cause a a change of the stellar abundance that is a factor of 2.5 times smaller compared to the case of 0.01 solar mass convective zone. In turn, Fig.~\ref{fig:convect} reveals that an ingested planet of 11.0 Earth masses in star with a convective zone  that is 2.5 times as large would cause the exact same abundance difference of the 4.4 Earth mass planet in the star with a convective zone of 0.01 solar masses. This is caused by the fact that the stellar convective zone is around $\sim$3-4 orders of magnitude larger than the planetary mass it ingests, and thus the relationship of eq.~\ref{eq:engulf} remains linear. The planet forming at 650 K shows similar results to the counterpart forming around a star with a 0.01 solar mass convective zone (Fig.~\ref{fig:pairs}).

However, the general trend in respect to which elements are influenced by the ingestion of the planet are unaffected. In the here presented example, Si, Fe, and Mg are influenced more compared to oxygen. This is caused by the formation temperature of the planet at 650 or 720 K. Consequently, our conclusions are affected by the mass of the stellar convective zone, which influences the mass of the material that the star has ingested, but the trend of the origin of the material, which sets the chemical abundance differences, remains independent of the stellar convective zone.

\textcolor{white}{\lipsum}

\end{appendix}

\end{document}